\documentstyle[aaspp4]{article}

\def \eg           {{e.g.}}
\def \etal         {{et~al. }}
\def \h2         {\hbox{H$_2$}}
\def \ie           {{i.e.}}
\def \IRAS         {\hbox{{\it IRAS\ }}}
\def \kms          {\hbox{km$\,$s$^{-1}$}}
\def\approxlt{\lower.2em\hbox{$\buildrel < \over \sim$}}
\def\approxgt{\lower.2em\hbox{$\buildrel > \over \sim$}}
\def \lco          {\hbox{$L_{\rm CO}$}}
\def \lhcn          {\hbox{$L_{\rm HCN}$}}
\def \ll            {\hbox{$\rm K \kms pc^2$}}
\def \lir          {\hbox{$L_{\rm IR}$}}
\def \ls           {\hbox{L$_{\odot}$}}

\def \ms           {\hbox{M$_{\odot}$}}           
\def \Msun         {\hbox{M$_{\odot}$}}           

\def \date         {\ifcase\month \message{zero} \or
                    January \or February \or March \or April \or May \or June 
                    \or July \or 
                    August \or September \or October \or November \or 
                    December \fi
                    \space\number\day, \number\year}

\begin{document}

\title{The Star Formation Rate and Dense Molecular Gas in Galaxies}
\author{Yu Gao$^{1,2,3}$ \  and \ Philip M. Solomon$^4$}
\affil{1 \ Purple Mountain Observatory, Chinese Academy of Sciences, 
2 West Beijing Road, Nanjing 210008, P.R. China}
\affil{2 \ Department of Astronomy, University of Massachusetts, \\
LGRT B-619E, 710 N. Pleasant St., Amherst, MA 01003}
\affil{3 \ Infrared Processing and Analysis Center, 
California Institute of Technology, MS 100-22, Pasadena, CA 91125}
\affil{4 \ Department of Physics \& Astronomy, SUNY at Stony Brook, 
Stony Brook, NY 11794}
\authoremail{gao@astro.umass.edu, psolomon@astro.sunysb.edu}


\begin{abstract}

HCN luminosity is a tracer of {\it dense} molecular gas, 
$n(H_2) \approxgt 3 \times 10^4$cm$^{-3}$, associated with 
star-forming giant molecular cloud (GMC) cores. We present the results 
and analysis of our survey of HCN emission from 65 infrared galaxies, 
including nine ultraluminous infrared galaxies (ULIGs,
\lir$\approxgt 10^{12}\ls$), 22 luminous infrared
galaxies  (LIGs, $ 10^{11}\ls< \lir \approxlt 10^{12}\ls$), and 34 
normal spiral galaxies with lower IR luminosity (most are 
large spiral galaxies). We have measured the global HCN line 
luminosity, and the observations are reported in Paper I. This paper 
analyzes the relationships between 
the total far-IR luminosity (a tracer of the star
formation rate), the global HCN line luminosity (a measure of the
total {\it dense} molecular gas content), and the CO luminosity 
(a measure of the total molecular content).
We find a  tight linear correlation between the IR and HCN luminosities
$L_{\rm IR}$ and $L_{\rm HCN}$ (in the log-log plot) with 
a correlation coefficient R = 0.94, and an almost constant 
average ratio $L_{\rm IR}/L_{\rm HCN} = 900 \ls/\ll$.
The IR--HCN linear correlation is valid  over 3 orders of magnitude 
including ULIGs, the most luminous objects in the local universe. 
The direct consequence of  the linear IR--HCN correlation is that the 
star formation law in terms of {\it dense} molecular gas content has 
a power law index of 1.0.
The global star formation rate  is linearly proportional to the mass of  
dense molecular gas in normal spiral galaxies, LIGs, and ULIGs. 
This is strong evidence in favor of 
star formation as the power source in ultraluminous
galaxies since the star formation in these  galaxies 
appears to be normal and expected given their high mass of dense 
star-forming molecular gas. 

The HCN--CO correlation is also much tighter than the IR--CO correlation.
We suggest that the nonlinear correlation between $L_{\rm IR}$ 
and $L_{\rm CO}$ may  be a consequence of the  stronger  and
perhaps more physical correlations between $L_{\rm IR}$ and 
$L_{\rm HCN}$ and between $L_{\rm HCN}$ and $L_{\rm CO}$. Thus, 
the star formation rate indicated by \lir~ depends on the amount of dense
molecular gas traced by HCN emission, not the total molecular gas traced 
by CO emission. One of the main arguments in favor of an active galactic
nucleus (AGN) as the 
power source in ULIGs is the anomalously 
high ratio \lir/\lco~ or \lir/M(H$_2$)  or high star formation
rate per \ms~ of gas, compared with that from normal spiral galaxies. 
This has been interpreted as indicating that a dust-enshrouded AGN 
is required to produce the very high luminosity. Viewed in 
terms of the dense gas mass the situation is completely different.
The ratio \lir/\lhcn~ or \lir/M$_{\rm dense}$, a measure of the star 
formation rate per solar mass of {\it dense} gas is essentially
the same in all galaxies including ULIGs. The ratio 
\lir/M$_{\rm dense}$  is virtually independent of
galaxy luminosity and on average \lir/M$_{\rm dense} \approx 90 \ls/\ms$,
about the same as in GMC cores but much higher than in GMCs.
We find that ULIGs simply have a large quantity of dense molecular
gas and thus produce a prodigious starburst that heats the dust, 
produces the IR, and blocks all or most optical radiation. 
The HCN global luminosity may be
used as an indicator of the star formation rate in high-redshift objects
including hyperluminous galaxies.
 
The HCN/CO ratio is an indicator of the dense molecular gas fraction.
and gauges the globally averaged molecular gas density. We find that 
the HCN/CO ratio is a powerful starburst indicator. All galaxies 
in our sample with a high dense gas
mass fraction indicated by $L_{\rm HCN}/L_{\rm CO} > 0.06$ are  
LIGs or ULIGs. Normal spirals all have similar and low 
dense gas fractions $L_{\rm HCN}/L_{\rm CO} = $ 0.02 to 0.05.
The global star formation efficiency depends on the fraction of 
the molecular gas in a dense phase.
 
\end{abstract}
\keywords{galaxies: ISM --- galaxies: starburst --- infrared: galaxies --- ISM: molecules --- radio lines: galaxies}


\section{INTRODUCTION}

Stars are born in the molecular interstellar medium,  the raw
material for star formation. In the Milky Way, all star  formation
essentially takes place in molecular clouds and  most star
formation takes place in giant molecular clouds (GMCs; Solomon,
Sanders, \& Scoville 1979) with mass M $> 10^5$
\Msun  and not the diffuse neutral  ISM dominated by  atomic
hydrogen (extended HI gas disk). The star formation rate (SFR) of
molecular clouds can be estimated from the far-infrared (FIR)
luminosity emitted by the warm dust heated by embedded high-mass 
OB stars (e.g., Mooney \& Solomon 1988). The mass of
molecular gas can be determined from the CO  luminosity calibrated
by $\gamma$  ray flux from the interaction of cosmic rays with
hydrogen molecules (\eg, Bloemen et al. 1986) or by dynamical cloud
masses determined from  CO kinematics for virialized individual
molecular clouds (Solomon et al. 1987; Young
\& Scoville 1991).  These methods are in good agreement (see
Solomon \& Barrett 1991). All strong  high-mass star formation regions are
associated with  GMCs, especially the cores of GMCs. The ratio of
FIR luminosity to the CO luminosity, or to the cloud mass, a
measure  of the SFR per solar mass of the cloud and an indicator of
star formation efficiency (SFE),  ranges over a factor  of 100 for
different clouds, and over a factor of 1000 from clouds  to the
cores of GMCs (e.g., Mooney \& Solomon 1988;  Plume
\etal 1997). An understanding of the physical conditions in GMCs
and their relation to galactic dynamics is a prerequisite to the
understanding of the star formation process, the SFR in galaxies
and starbursts.

Star formation in galaxies is closely tied up with the local gas 
density, as formulated in the Schmidt (1959) law, although the
important component is the molecular gas.  Globally, the SFR 
correlates with the molecular gas content in galaxies, as traced by
CO emission, including luminous and ultraluminous infrared galaxies
(LIGs and ULIGs\footnote{LIGs: 
$10^{12}\ls \approxgt \lir >10^{11}\ls$,  ULIGs: \lir$\approxgt
10^{12}\ls$ (to be  exact, 10$^{11.9}$\ls~ in this paper). For a
definition of the total IR (8 to 1000$\mu$m) luminosity \lir~  and
FIR luminosity $L_{\rm FIR}$, see Sanders \& Mirabel  (1996).
The value of $L_{\rm IR}$ is generally larger, by up to
$\sim$20\% as it includes both  12 and 25$\mu$m emission, than
$L_{\rm FIR}$. However, we often simply refer  the total IR
emission as FIR in this paper.}).
 In Galactic star-forming regions,  active high-mass star formation
is intimately related to the very dense  molecular gas in the cores. 
While the canonical molecular  gas tracer CO   shows strong
emission in cloud cores, it is  not specific enough to reveal their
star formation potential.   The bulk of the cloud material traced by
CO observations is  in the GMC envelopes and is at a much lower
density.  The physical conditions of active star-forming GMC cores 
are better revealed by emission from very high transition CO lines
(in the submillimeter regime) and  high dipole-moment molecules 
like CS and HCN.

HCN is one of the most abundant high  dipole-moment molecules
that traces molecular gas at densities  $ n(H_2) \approxgt
3\times 10^4$cm$^{-3}$, more than 2 orders of magnitude 
higher than that traced by CO ($\approxgt 300$cm$^{-3}$).  Many
HCN (1-0) observations have already been conducted by different
groups (Nguyen-Q-Rieu \etal 1989, 1992;  Henkel \etal 1990;
Solomon, Downes, \& Radford 1992;  Israel 1992; Helfer \& Blitz
1993; Aalto \etal 1995; Curran, Aalto, \& Booth 2000) in a variety  of
nearby galaxies. These previous  observations contain small
samples with frequent overlap in the sample selection. The total
number  of galaxies detected in HCN is still small ($\approxlt 30$),
and for many galaxies only the central position was observed. Gao
\& Solomon (2003, hereafter Paper I), in the companion paper,  have
presented  a systematic  survey  of global HCN luminosity that 
more than doubles the number of  galaxies ($\sim 60$) observed in
HCN.

Although the dense molecular gas is strongly concentrated in the
central  regions ($\approxlt 1$\,kpc),  HCN mapping of a dozen
nearby galaxies out to  a diameter of $\sim D_{25}/4$ (Gao 1996,
1997; Y. Gao \& P.M. Solomon 2004, in  preparation) shows that a substantial
fraction of  HCN emission originates from the inner disks  outside
the central $\sim 1$~kpc.  All previous HCN observations in external
galaxies including a few HCN maps (e.g., Nguyen-Q-Rieu et al. 1992;
Reynaud \& Downes 1997; Helfer \& Blitz 1997a)  have primarily
been observations of the galactic central regions.  In most cases,
the total HCN emission from the entire galaxy  has not been 
measured. The recent HCN observations in 20 Seyfert galaxies are 
primarily of  the galaxy centers, and many are nondetections or
marginal detections (Curran \etal 2000).  Confusion may occur when
the results drawn from the HCN observations of central regions of
galaxies  are compared with the global properties of galaxies. Aalto
\etal (1995)  have not found a correlation between the molecular
line intensity ratio, $I_{\rm CO}/I_{\rm HCN}$, and FIR emission, or
measures of star-forming activity in their sample of 10 interacting
galaxies, which appeared to be in conflict with  the original
findings of a tight FIR--HCN correlation  in another sample of 10
LIGs/ULIGs and spiral galaxies where the total HCN emission was
measured (Solomon \etal 1992). The situation has now been clarified
by our new HCN survey of $\sim 60$ galaxies (Paper I),  which 
measured the global HCN
emission in a wide range of galaxies.   There is indeed a tight
FIR--HCN correlation in a statistically significant HCN sample.

It is clear from all previous observations that  the molecular gas in
the central regions of spiral  galaxies, starbursts, and LIGs/ULIGs is
much denser than  the molecular ISM in the disks of spiral galaxies
(\eg, Nguyen-Q-Rieu \etal 1992; Solomon \etal 1992; Helfer \& Blitz
1997a; Wild \& Eckart 2000). From various observational studies of
star-forming regions of GMCs in the Milky Way,  it is also clearly
shown that the active star-forming regions in the disk are the
dense molecular cloud cores (\eg, Mooney \& Solomon 1988; Plume
\etal 1997; Pirogov 1999; Evans 1999) rather than the entire
molecular envelopes of GMCs.  The SFE of active star-forming
clouds that are associated with  IR sources readily apparent on
\IRAS 60 and 100 $\mu$m images can be 100 times higher than the
IR-quiet clouds with no apparent IR sources revealed by \IRAS
(Mooney \& Solomon 1988). Star formation efficiency 
($L_{\rm IR}/L_{\rm CO}$) can vary
over a factor of 100 as well  in galaxies. 

Normal spiral galaxies have an SFE similar to that of Galactic GMCs,
whereas the SFE of ULIGs/LIGs and advanced mergers can be more
than an order of magnitude larger (\eg, Solomon \& Sage 1988;
Solomon \etal 1997; Gao \& Solomon 1999).  These differences in SFE
can be understood in terms of the different dense molecular gas
content as traced by HCN observations.

For 10 galaxies, including both  LIGs/ULIGs and normal spiral
galaxies, Solomon \etal (1992) show that there is a tight correlation
between the ratio of the IR and CO luminosities $L_{\rm IR}/L_{\rm
CO}$  and the HCN/CO luminosity ratio $L_{\rm HCN}/L_{\rm CO}$  in
addition to the excellent correlation between \lir~ and \lhcn. This is
now fully confirmed with our current HCN study of a sample of 65
galaxies (\S 3).
 These correlations demonstrate a close  relationship between the 
SFR and  the {\it dense molecular gas reservoir} in galaxies. The SFE
depends on the fraction of available molecular gas in a dense phase 
($L_{\rm HCN}/L_{\rm CO}$), and the dense molecular content of
even gas-rich spirals is much less than that of LIGs/ULIGs of
comparable total molecular gas content  (Solomon \etal 1992;
Radford 1994). Since CO emission traces most of the molecular gas
mass and is not necessarily a specific tracer of the {\it dense}
molecular gas (e.g., Mauersberger \& Henkel 1993; Evans 1999) or 
the IR luminosity from star formation (Mooney \& Solomon 1988), 
CO alone can give a misleading picture of the densest molecular 
gas in a galaxy. 

In Galactic plane GMCs, for example, essentially all OB
star formation occurs in the cores of GMCs with strong CS and HCN
emission.  The ratios of CO/CS and CO/HCN intensities for Galactic
disk GMCs are much larger than for Galactic center clouds (Lee, 
Snell, \& Dickman 1990; Jackson \etal 1996; Plume \etal 1997; 
Helfer \& Blitz 1997b) and an order of magnitude larger than for the
archetypal ULIG  Arp~220 (Solomon, Radford, \& Downes 1990). In
some galaxies, the molecular gas in the center is much more
prominent in HCN emission than in CO.  A good example is the center
of the Seyfert 2/starburst hybrid galaxy NGC\,1068, where all 
interferometric maps  demonstrate that the nuclear region is more
prominent than the  rings or spiral arms when viewed in HCN, while
the opposite is true in CO emission (Tacconi \etal 1997, 1994;  Helfer
\& Blitz 1995; Jackson \etal 1993). A similar trend is also observed in
the centers of M51 and NGC 1097 (\eg, Kohno \etal 1996, 2003).

Here we utilize a large, statistically significant sample with
observations of global HCN emission  from  65 spiral galaxies, LIGs,
and  ULIGs  --- 53 from the systematic  HCN survey of Paper I, 10
from  Solomon, Downes, \& Radford (1992), plus two from the
literature. These galaxies range over 3 orders of magnitude in FIR
luminosity.  We analyze the various relationships among the global
HCN, CO, and FIR luminosities.   We further discuss the physical
relationship between  the dense molecular gas content and the rate
of high-mass star formation in galaxies. 

The HCN sample and observations are briefly reviewed in \S 2. 
Section 3 presents the results. Section 3.1 is a comparison of the IR--HCN and
IR--CO correlations. Section 3.2  concentrates on the importance of the
dense gas mass fraction and the star formation efficiency.  Section 3.3
and the Appendices  present multi-parameter fits to the data and the
effect of dust temperature. Section 3.4 has   a brief summary  of all
results.  In \S\S 4.1--4.3, we discuss the  importance of HCN as a
tracer of star-forming molecular gas, the star formation rate as a
function of the dense gas mass, and the global star formation law. 
Section 4.4 discusses the HCN/CO ratio as a starburst indicator.  Section 4.5
discusses the origin of FIR emission from spirals and ULIGs. Section 4.6
briefly speculates on the implications for hyperluminous infrared
galaxies at high $z$. Finally, we summarize the main points of our
study in \S 5.

\section{THE HCN SURVEY} 

The detailed descriptions of our survey sample and HCN (and CO)
observations were given in the companion paper (Paper I). 
Our HCN survey sample is drawn from samples in recent CO 
observations of galaxies showing strong CO emission (\eg,
the CO antenna brightness temperature much larger than 
100 mK for normal spiral galaxies and larger than 20 mK
for LIGs/ULIGs). All truly
IR-bright galaxies with 60 or 100$\mu$m emission larger
than 50 or 100 Jy, respectively, have also been included.
Essentially all galaxies with strong CO and IR emission in northern
sky have been chosen for the HCN survey. 

We carried out 
several observing runs mainly with the former NRAO\footnote{The 
National Radio Astronomy
Observatory is operated by Associated Universities, Inc., under
cooperative agreement with the National Science Foundation.} 12m
telescope at Kitt Peak, the IRAM 30m telescope at Pico Veleta
near Granada, Spain, and the FCRAO 14m telescope for  most of 
our observations. We here only
mention our observing strategy of using different telescopes.
The IRAM 30m was mostly used to observe ULIGs (rather distant)
and to map out some nearby starburst galaxies of smaller optical
diameters, given the small matching beam size as compared with 
the source extent for effective mapping. 
Essentially all other observations were conducted with 
the NRAO 12m so that one beam measurement can cover almost all HCN
emission from relatively distant galaxies, including some merging
galaxy pairs. Also almost all nearby large spiral galaxies have 
been at least mapped along the major axes with the NRAO 12m, and only 
a few of them were initially tried mapping with the
QUARRY receivers at the FCRAO 14m 
(because of its limited sensitivity). 

The goal of the HCN survey of Paper I
is to determine the total HCN emission from the whole inner 
disks or the entire galaxies in a large sample of galaxies 
with a wide range of IR luminosity. 
Combined with the HCN data of 12 galaxies, 10 from 
Solomon \etal (1992), including a half-dozen
LIGs/ULIGs, and the other two galaxies, M51 
(Nguyen-Q-Rieu \etal 1992) and NGC~4945 (Henkel, 
Whiteoak, \& Mauersberger 1994) that were mapped extensively in HCN, 
we have a large, statistically significant sample 
of 65 galaxies with globally measured HCN emission. 
In addition, there are more than a dozen nearby large spiral
galaxies with HCN detections toward the galactic nuclei 
(Nguyen-Q-Rieu \etal 1992; 
Helfer \& Blitz 1993; Aalto \etal 1995; Curran \etal 2000), 
including LIGs (\eg, NGC~3256, Casoli, Dupraz \& 
Combes 1992) that 
do not overlap with our HCN survey sample\footnote{However, 
Curran et al. (2001) submitted and published several HCN maps 
(though mostly still very limited spatial coverage of only central 
1$'$--2$'$ regions) of nearby Seyfert galaxies 
after we submitted our paper.}. In
principle, we could further enlarge our HCN sample to
a total of about 80 galaxies. These observations of nearby
galaxies did not measure the total HCN emission, however,
with only the central/nuclear HCN emission observed. 
Therefore, most HCN data in these nearby galaxies can thus only 
be used to set 
lower limits to the total HCN luminosities. Further HCN 
observations at least along the major axes are required to map out
the total HCN emission. For the sake of completeness, consistency, 
and uniformity of our sample, these galaxies with lower HCN limits 
in the literature are not included in our sample.

The derived global properties of the line luminosities and various
luminosity ratios of galaxies in the HCN survey sample are listed 
in Table~1 (cf. Table~4 in Paper I), 
together with a dozen galaxies where the 
total HCN luminosities are available or can be estimated 
from the literature. 
There are also several other fairly distant galaxies in Curran et al. (2000)
and Aalto et al. (1995) in which the global HCN luminosities were 
supposedly measured, but we did not include them in our sample
as their data seem to have low signal-to-noise ratio. 
Mrk\,273, one of the galaxies that overlaps in both
of our sample and that of Curran et al. (2000), for example, was 
claimed to be detected with an extremely large HCN/CO luminosity 
ratio of 1. Our high-quality IRAM 30m spectra (Paper I), 
however, clearly indicate an HCN/CO intensity ratio of $\sim 1/6$ and 
a HCN/CO luminosity ratio of 0.23, which is consistent with all other ULIGs 
that have been significantly detected in HCN so far, rather than 
an unrealistic ratio of $\sim 1$. 

We also have some limits to the total HCN line 
emission owing to insufficient mapping of HCN (about 10~\% of the 
galaxies in our sample), but our HCN lower limits are probably
close to the true values as we have some
off-nucleus HCN measurements besides the central
beams on the nuclei in most cases. We keep these sensitive HCN limits in 
Table~1 and include these data for the various analyses.

\section{RESULTS AND ANALYSIS}

\subsection{Comparison of the IR--HCN and IR--CO Correlations}

The principal observational result from this survey is  
the tight linear relation observed between
far-infrared luminosity, $ L_{\rm IR}$, and HCN line luminosity, 
$ L_{\rm HCN}$, shown in  Figure~1a. This very good correlation 
extends over 3 orders of magnitude in luminosity and includes 
normal spiral galaxies and luminous and ultraluminous infrared galaxies 
(LIGs/ULIGs).  
Figure~1b  shows the correlation between  $L_{\rm IR}$  and  $L_{\rm CO}$
of the same sample, which has a larger scatter than the FIR--HCN  
relation and most importantly
steepens with higher IR luminosity, showing the well-known result 
that LIGs/ULIGs, although rich in molecular gas, have a substantially 
higher IR luminosity per unit CO luminosity or per solar mass of 
molecular gas \h2  (\eg, Solomon  et al. 1997).

A least--squares fit using all the data but excluding HCN limits
 yields a power-law slope of
 1.00 $\pm$ 0.05 and $1.25 \pm 0.08$ 
for \lir--\lhcn~   and  \lir--\lco,  respectively. The corresponding
correlation coefficients (squared) are $R^2$ =  0.88   and 0.77.
The slopes change  to  1.05 $\pm$ 0.05 and $1.44 \pm 0.08$ 
for \lir--\lhcn~   and  \lir--\lco, respectively, 
if an orthogonal fit is used.
Including the sources with HCN limits in
the fit has little effect on the slope or the fit.
The best-fit (logarithmic) relation between HCN and IR luminosities
excluding galaxies with upper or lower limits   is
\begin{equation}
logL_{\rm IR}=1.00 (\pm 0.05) logL_{\rm HCN}+2.9, ~or~~
\lir/\lhcn=900\ls/\ll,
\end{equation}
indeed a linear relation. When we use all galaxies
including HCN limits the correlation remains almost the same
with \ $logL_{\rm IR}=0.97logL_{\rm HCN}+3.1$ \
and with the same correlation coefficient (R=0.94).


At first glance of Figure~1, the correlation between
$L_{\rm CO}$ and  $L_{\rm IR}$ may  appear to be nearly as good as that
of the $L_{\rm HCN}$--$L_{\rm IR}$ correlation, even though there
is significant difference in the correlation coefficients and 
the dispersion from the fit. The most obvious difference occurs at the
high $L_{\rm IR}$ end. In Figure~1b, a line of constant 
$L_{\rm IR}$/$L_{\rm CO}=33\ls/\ll$ fits the lower luminosity
galaxies but lies well below almost all (28/31) of the high-luminosity 
galaxies (LIGs/ULIGs). The correlation between
$L_{\rm IR}$  and  $L_{\rm HCN}$ still fits the ratio   determined from 
 lower IR luminosities ($L_{\rm IR}<10^{11} \ls$), but
the correlation between $L_{\rm CO}$ and $L_{\rm IR}$ does not.
(The lines shown in Fig.~1 are the fits of a fixed slope of 1
to low-luminosity galaxies.) The FIR--HCN relation is linear all the way 
up to $L_{\rm IR} = 10^{12.5} \ls$, but the FIR/CO  ratio is 
systematically higher  for high-luminosity galaxies than for those of
normal luminosity.  

One obvious difference between the FIR--CO and FIR--HCN
relationships is that the range of CO luminosity is only 2 
orders of magnitude, whereas HCN luminosity covers 3 orders of 
magnitude, almost the same as the IR luminosity.
For LIGs/ULIGs the CO is systematically weak compared with the IR.
  Measurements of  CO luminosity of a large sample
of ULIGs (Solomon \etal 1997; most of which are not in Fig. 1b since there are
no corresponding HCN observations  available) show that they all lie above the
fitted line in Figure\,1b.  


Another way to demonstrate these points is to compare the luminosity
ratios of $L_{\rm IR}/L_{\rm HCN}$ and $L_{\rm IR}/L_{\rm CO}$ with
$L_{\rm IR}$ (Fig. 2) . 
 $L_{\rm IR}/L_{\rm HCN}$ appears to be nearly
independent   of $L_{\rm IR}$, whereas the $L_{\rm IR}/L_{\rm CO}$
ratio increases substantially  with increasing $L_{\rm IR}$ as
established by previous work (\eg, Young \etal 1989; Sanders,
Scoville, \& Soifer 1991). 
These results are also revealed in Table 2, which summarizes 
the average IR, CO, and HCN luminosities
among ULIGs, LIGs, normal spiral galaxies, and all galaxies
of the entire HCN sample.


In summary, the IR--HCN correlation is linear and is extremely tight 
over 3 orders of magnitude in luminosity, when compared with the 
nonlinear IR--CO correlation. While the high  luminosity of LIGs/ULIGs 
requires an  elevated SFE of the total molecular gas indicated by 
$L_{\rm IR}/L_{\rm CO}$, the SFR per unit of {\it dense} molecular 
gas---the SFE of the {\it dense} molecular gas indicated 
by ($L_{\rm IR}/L_{\rm HCN}$)---is  almost constant 
and independent of the IR luminosity or total SFR. The
tight relationship between $L_{\rm IR}$ and $L_{\rm HCN}$ will be 
further illustrated by comparing it with other correlations in the 
following two sections (\S3.2 \& \S3.3).

\subsection{Dense Molecular Gas Fraction (\lhcn/\lco) and Star Formation Efficiency (\lir/\lco)} 

Figure 3 shows a significant correlation between HCN and CO
luminosities with a correlation coefficient R=0.92 ($R^2=0.85$).  
In all figures, except where
$L_{\rm IR}$ is itself explicitly plotted, we distinguish LIGs/ULIGs 
($L_{\rm IR}>10^{11} \ls$, filled circles) from the less luminous 
normal spiral galaxies (open circles) so that any systematic 
difference between these two subsamples can easily be seen.  
This suggests that the more gas-rich
galaxies tend to have more dense molecular gas (and vice versa),
and thus are  more luminous (Fig.~1).  The correlation is
much better for normal galaxies (open circles)  than for LIGs/ULIGs 
(filled circles). For normal spiral galaxies  there is a tight  
relationship  between  the HCN and CO luminosities with a slope
of 1.0, shown as the fit line in Figure 3. There is also a very small 
dispersion about the fit with  $\sigma (log L_{\rm HCN} ) = 0.14$.
 This good linear HCN--CO correlation is 
 the underlying reason that the FIR--CO correlation
is linear for normal galaxies and CO luminosity  is reasonably 
good at predicting their SFR (FIR). 

 All  but three of the  31 LIGs have  $L_{\rm HCN}$ above the linear fit 
(Fig. 3) determined from normal spirals, clearly showing that excess HCN
emission compared with CO emission is a  characteristic of LIGs/ULIGs. An
orthogonal fit to all galaxies gives a nonlinear relation:
\begin{equation}
logL_{\rm HCN}=1.38 logL_{\rm CO}-4.79, ~or~~ 
{\lhcn\over{\lco}}=0.1({\lco\over{10^{10}\ll}})^{0.38}. 
\end{equation}
It is the high ratio of HCN/CO for LIGs/ULIGs that causes the breakdown 
in the IR--CO relation. 

A high ratio of  $L_{\rm HCN}/L_{\rm CO}$  is a distinguishing feature 
of LIGs/ULIGs. As can be seen from Table 2, the average ratio systematically 
increases as IR luminosity increases.  Figure~4 shows the IR luminosity 
of all 65 galaxies in the sample as a function of the ratio
$L_{\rm HCN}/L_{\rm CO}$. It is immediately apparent that LIGs/ULIGs
($L_{\rm IR}>10^{11}\ls$) have systematically higher HCN/CO ratios 
than normal galaxies.  However, for normal spirals only, although there 
is some scatter with $L_{\rm HCN}/L_{\rm CO} = 0.02$--0.06, there
is no correlation between
IR and \lhcn/\lco.  All of the most luminous galaxies, \ie, ULIGs, have high 
$L_{\rm HCN}/L_{\rm CO} > 0.09$.    The most striking result is that 
{\it all ( 27/27)  galaxies with  $L_{\rm HCN}/L_{\rm CO} >  0.06$ 
are luminous (or ultraluminous)}. 
The 2 most luminous galaxies in our sample Mrk~231 and IRAS~17208-0014 
have HCN/CO = 0.23 and 0.25, respectively.  The molecular ratio 
$L_{\rm HCN}/L_{\rm CO}$ appears to be $100\%$ successful as an 
indicator of  galaxies with infrared starbursts.  
In \S~4 we further discuss the contributions of the disk and nuclear 
regions to these global values of \lhcn~ and \lco~ as well as the
\lhcn/\lco~ ratio.

While all galaxies with a high HCN/CO ratio are IR-luminous, it 
is not true that all LIGs have a high dense gas fraction,
$L_{\rm HCN}/L_{\rm CO}$. There are a group of seven LIGs in our 
sample that are IR
luminous and have high HCN luminosities but normal ratios of HCN/CO. 
These are gas-rich (CO-luminous)
galaxies with normal SFE. They are the filled circles on the normal 
galaxy fit line in Figure~3 and in the
upper left of Figure 4.  Their high
\lir\ is  simply  the result of a  tremendous amount of available 
molecular gas and a normal
fraction of dense molecular gas.  In this sense they are not starbursts 
globally. They are globally using their
molecular gas at a normal SFR.  Some of the notable examples in our sample are
Mrk~1027, NGC~1144 (Gao \etal 1997), NGC~6701, and even Arp~55 (Table~1).
Luminous prestarburst galaxies like Arp~302 (Gao 1996; 
Lo, Gao, \& Gruendl 1997), and even early stage galaxy mergers 
in the initial starburst phase with extended CO emission
(\eg, NGC~6670, Wang \etal 2001), belong to this category as well.  


Figure 5a shows a correlation between $L_{\rm IR}/L_{\rm CO}$
and $L_{\rm HCN}/L_{\rm CO}$ with a correlation coefficient
$R=0.74$ ($R^2=0.55$, there is little
effect on the fit whether those HCN limits are used or not). 
This suggests that the star formation efficiency, 
SFE ($L_{\rm IR}/L_{\rm CO}$), 
depends on the fraction of molecular gas in a dense phase indicated by 
($L_{\rm HCN}/L_{\rm CO}$).
This correlation demonstrates the direct connection between 
the IR and HCN luminosities (as shown in Fig.~1a).  Both 
\lir~ and ~\lhcn~ have been normalized by \lco\, to show
the physical relationship between \lir~ and \lhcn~ after removing the
dependence upon distance, galaxy size, and  
other possible selection effects.  Of course, the range of the ratio  
($L_{\rm HCN}/L_{\rm CO}$) is less
than that for  $L_{\rm HCN}$ and the correlation is  not as strong 
as in Figure 1a, although the
dispersion about the fit is almost the same. 
This normalization is dust extinction--free, unlike
the use of, for example, the blue luminosity in the normalization.


Similarly, we can show the correlation between $L_{\rm IR}$ and 
$L_{\rm CO}$ divided by ~\lhcn~ for normalization. Surprisingly, 
the strong correlation observed in Figure~1b has completely disappeared,
and there is no correlation left at all ($R^2=0.01$; Fig.~5b). 
This certainly reflects the tight correlation between IR and HCN 
and the fact that  $L_{\rm IR}/L_{\rm HCN}$  is almost   
independent of galaxy luminosity (Fig.~2a). 
The absence of a correlation in Figure~5b also implies that 
the apparent well-known correlation between 
$L_{\rm IR}$ and $L_{\rm CO}$ may lack a true physical basis, 
at least as compared with that
between $L_{\rm IR}$ and $L_{\rm HCN}$. 
 The IR--CO correlation may be due to a combination of the 
 tighter IR--HCN and HCN--CO correlations. 
Nevertheless, $L_{\rm IR}/L_{\rm HCN}$ is not completely independent 
of the CO luminosity (see \S 3.3 and the appendix) as
its inverse $L_{\rm HCN}/L_{\rm IR}$  is weakly correlated with  
$L_{\rm CO}/L_{\rm IR}$ with a correlation coefficient $R=0.56$ 
($R^2=0.31$, still a meaningful correlation as a consequence of
the tight HCN--CO correlation normalized by \lir). 

\subsection{The Model Parameter Fits}

Table~3 in Appendix B summarizes almost all the parameter fits 
in the IR, CO, and HCN data sets including the warm dust temperature 
deduced from the 60 and 100 $\mu$m fluxes. Here we discuss 
the three-parameter fits involving the IR, HCN, and CO luminosities
and the warm dust temperature $T_{\rm dust}$ dependence of the 
luminosity ratios. Further details of various 
other correlation fits are also discussed in the Appendix. 
These fits to the data demonstrate that the two important 
independent parameters are the IR and HCN luminosities.

\subsubsection{Multi-parameter Fits}

 The  IR luminosity from a model fit to the HCN and 
CO [the IR(\lhcn, \lco) model] yields 
\begin {equation}
logL_{\rm IR}(L_{\rm HCN},
L_{\rm CO})=2.28+0.88logL_{\rm HCN}+0.16logL_{\rm CO}. 
\end {equation}
The weak dependence on \lco~ shows that the IR luminosity is determined 
principally from the HCN luminosity. The dispersion about the fit is 
almost the same as the two parameter fit of \lir--\lhcn.
 The t ratio for \lco~ (0.98) is very  small
as compared with that for ~\lhcn~ (6.95), demonstrating  that \lco~ is almost
random after fitting ~\lhcn~ for \lir. 

An IR and CO luminosity model fit for \lhcn~ [the HCN(\lir, \lco) model]
gives 
\begin {equation} logL_{\rm HCN}(L_{\rm IR},~L_{\rm CO})=
-2.62+0.56logL_{\rm IR}+0.49logL_{\rm CO}.
\end {equation}
 Here, \lir~  and  \lco~ seem to be almost equally important.  
This relation produces a tighter fit
than either of the two-parameter fits  HCN--CO (Fig. 3) or HCN--IR (Fig. 1a). 

Therefore, although  the HCN luminosity depends on both  the CO and IR  
luminosity, the IR  is predicted  almost entirely by the  HCN with 
almost no effect from the  CO luminosity. The reason for this 
surprising result is probably that, 
of the three separate two-parameter (IR--HCN, HCN--CO, and IR--CO)
fits,  the weakest and the one that has the largest dispersion of 
the data from  the fit is the IR--CO correlation. 

The corresponding fit for the CO luminosity is
\begin {equation}
 logL_{\rm CO}(L_{\rm IR},~L_{\rm HCN})=
3.60 +0.55logL_{\rm HCN}+0.12logL_{\rm IR}.
\end {equation}
It is clear that the CO is more closely related to the HCN than 
to the IR  luminosity. In other words, the dependence of the 
CO luminosity on the IR is much 
weaker than on the HCN luminosity. This result is probably due to 
the very tight correlation between CO and HCN for normal
galaxies (Fig. 3).  

\subsubsection{Warm Dust Temperature Dependence}

Here we examine the influence of the warm dust temperature on the 
luminosity ratios. It is well known that the warm dust temperature 
is an important parameter in the IR--CO correlation 
(\eg, Young \etal 1986; Young \& scoville 1991). Calculating 
the warm dust temperature from the 60/100 $\mu$m flux ratio, 
we find that the $L_{\rm IR}/L_{\rm CO}$ luminosity ratio depends 
strongly on $T_{\rm dust}$ (see figures in the appendix)
\begin{equation}
L_{\rm IR}/L_{\rm CO} \ = \ 61(T_{\rm dust}/35{\rm K})^{5.7}, 
\end{equation}
for optically thin dust with emissivity $\sim \nu^\beta$ ($\beta=1.5$).

 As we can see from Figure 2a, the ratio $L_{\rm IR}/L_{\rm HCN}$ is 
almost independent of IR luminosity. If we fit this ratio to 
the dust temperature (poor correlation), we find a weak correlation 
with 
\begin{equation}
L_{\rm IR}/L_{\rm HCN} \ \sim \ 980(T_{\rm dust}/35{\rm K})^{1.8}. 
\end{equation}

The IR--HCN relation changes only by a factor of 2 on average across 
the entire temperature
range of the sample galaxies (T = 29 to 46K), while the IR--CO ratio 
changes by a factor of 13. The HCN/CO ratio changes with the dust
temperature between these two extremes since (using eqs. [6] and [7])
\begin{equation}
L_{\rm HCN}/L_{\rm CO} = (L_{\rm IR}/L_{\rm CO})/(L_{\rm IR}/L_{\rm HCN})
\sim T_{\rm dust}^{3.9}.
\end{equation}
Indeed, a direct fit to the data (still a reasonable
correlation with $R^2 \sim 0.5$) gives 
$L_{\rm HCN}/L_{\rm CO} \sim T_{\rm dust}^{3.6}$.

\subsection{Summary of the Results}

Our results presented in Figures 1--5 show that dense molecular gas 
mass indicated by HCN luminosity is a much better predictor of 
infrared luminosity and star formation than   total H$_2$
content indicated by CO luminosity. We  identify two types of LIGs  
according to their total molecular gas content. Both types have high total
$L_{\rm HCN}$ in addition to  high $L_{\rm IR}$. Most LIGs have a high 
dense molecular gas fraction, but a second group has 
a normal ratio of HCN/CO yet a very high CO luminosity. 
Are both types of LIGs starbursts? It may well be that LIGs with 
a normal $L_{\rm HCN}/L_{\rm CO}$ are not necessarily  genuine
starbursts since the infrared luminosity can be produced by a large 
amount of molecular gas forming stars at a normal rate (efficiency). 
Star formation efficiency can remain almost unchanged, whether 
galaxies are molecular gas-rich (more luminous) or relatively 
gas-poor (less luminous, cf. Young 1999).
However, SFE increases dramatically from {\it dense}
molecular  gas-poor galaxies to {\it dense} molecular gas-rich 
galaxies. All galaxies with a dense
gas mass fraction  $L_{\rm HCN}/L_{\rm CO} > 0.06$ are LIGs or
ULIGs in our sample. In essence, HCN traces the molecular gas at high 
density and at high warm dust temperature that is tightly linked 
to the active star formation.

\section{DISCUSSION}

\subsection{HCN --- Tracer of Active Star-forming Dense Molecular Gas}

All the parameter fits essentially show the same thing.
CO luminosity by itself leads to a rough 
prediction for IR luminosity that breaks down for luminous infrared
galaxies (LIGs), especially for ultraluminous infrared
galaxies (ULIGs), whereas HCN luminosity is much 
better at predicting the IR luminosity for all galaxies including
ULIGs. Therefore, the star formation rate (SFR) indicated by \lir~ 
in galaxies 
depends on the amount of dense molecular gas traced by HCN, not the 
total molecular gas content measured by CO. In particular, the IR--CO 
correlation may not have a solid physical basis as it can 
be readily related to the stronger and perhaps more physical 
IR--HCN and HCN--CO correlations, which may be the origin of the 
correlation between IR and CO. This is reminiscent of the poor 
IR--HI correlation as compared with the better IR--CO correlation 
that became apparent two decades ago, when systematic CO observations 
of significant numbers of galaxies became available. 

The HCN radiation is associated with the warm dust traced by 
the FIR radiation, whereas the total
molecular gas traced by CO originates in diverse dust
components at different temperatures. 
The temperature dependence of \lir~ (or $L_{60 \mu m}$ or 
$L_{100 \mu m}$) is straight-forward to understand
since it is just the Planck law plus the dust emissivity. In this
perspective, it is easy to see why the \lir--\lco~ relation 
has a strong dependence upon the dust temperature term since
\lco~ at most is proportional to the first power
of the temperature, while \lir~ (or $L_{100 \mu m}$), depending on
the emissivity law, is proportional to at least the fifth power of 
the dust temperature. Therefore, many correlations that
involve $T_{\rm dust}$ and \lco~ [see Appendix, \eg,
in one case as represented in the 
CO($L_{100 \mu m}$, $T_{\rm dust}$) model, 
$log L_{\rm CO}=5.07+0.88logL_{100 \mu m}-3.0logT_{\rm dust}$]
can be easily explained. The more complex question is why this 
$T_{\rm dust}$ dependence almost entirely goes away for 
the \lhcn, \eg, the HCN($L_{100 \mu m}$, $T_{\rm dust}$) model,
$log L_{\rm HCN}=-1.12+0.95logL_{100 \mu m}-0.2logT_{\rm dust}$,
and similarly others involving \lhcn~ (Table~3). The simple 
answer to this is that the higher molecular gas density produces 
more active star formation, which raises 
$T_{\rm dust}$ owing to heating of the newborn stars. 
High molecular gas density, strong HCN emission, and warm 
dust temperature go together. HCN traces the active star-forming 
molecular gas where both the molecular gas density and dust 
temperature are high. 

We have already shown in Paper I that HCN emission
in galaxies is primarily due to the collisional excitation by
high-density molecular hydrogen, not radiative excitation through 
the mid-IR pumping (see also Stutzki \etal 1988; Paglione \etal 1997). 
Even though the mid-IR pumping is not
a significant source to excite the rotational transition of HCN emission, 
there are still some other possibilities that may help excite HCN, 
\eg, collisions with electrons (Aalto \etal 1995), the possibly enhanced 
HCN abundance and shock excitation, owing to excess supernovae occurred 
in starburst galaxies and LIGs/ULIGs. However, all these, even 
collectively, are only a secondary effect on a global scale in
galaxies, although significant contribution in a particularly favorable 
environment in some small localized regions cannot be excluded. 

In any case, the physical explanation for the tight correlation between
the HCN and IR is star formation in dense molecular gas. The active
high-mass star-forming sites are the cores of GMCs,
where the molecular gas is warmer and denser than in the GMC envelopes,
where most of the CO emission originates.
Currently, detailed statistical study examining the relationships 
among FIR, HCN, and CO on the scale of GMCs cores is not yet 
available. There are extensive observations (Plume \etal 1997)
of another dense gas tracer, CS emission, of high-mass star formation 
cores in the Milky Way. All of these regions have an H$_2$ density more 
than sufficient to produce strong HCN emission.

\subsection{Dense Molecular Gas and Star Formation Rates}
 
For an initial mass function (IMF), typically taken to be
the Salpeter IMF, $L_{\rm IR}/M$(gas) can be 
interpreted as a measure of star formation efficiency (SFE), 
\ie, SFR per unit gas mass. 
This is because the SFR is related to \lir~ by
\begin{equation}
\dot M_{\rm SFR} \approx 
2\times 10^{-10} (L_{\rm IR}/\ls) \ \ms  yr^{-1}, 
\end{equation}
assuming that the observed FIR emission is produced primarily 
from dust heating by O, B, and A stars (\eg, Scoville \& Young 1983; 
cf. Gallagher \& Hunter 1987; Kennicutt 1998b). 
Although \lir~ correlates with \lco, the correlation is nonlinear,
with a higher \lir/\lco~ ratio for higher \lir~ (Fig.~2b). On 
the other hand, \lir~ linearly correlates with ~\lhcn, implying an almost 
constant SFR per unit of {\it dense} molecular gas mass for all galaxies. 

The HCN luminosity can be related to the mass of dense gas,
$M_{\rm dense}=\alpha_{\rm HCN} \lhcn$, if we assume the
emission originates in the gravitationally bound cloud cores 
(see Paper I). For a volume-averaged
core density $n({\rm H_2}) \sim 3\times 10^4$cm$^{-3}$ and
brightness temperature $T_b=35$~K, 
$\alpha_{\rm HCN}=2.1 \sqrt{n({\rm H_2})}/T_b=10 \ms/\ll$.
Substituting in equation (1) gives a luminosity to
dense gas mass ratio
\begin{equation}
\lir/M_{\rm dense}=90(\alpha_{\rm HCN}/10) \ls/\ms
\end{equation}
for all galaxies, although the mean is actually slightly
higher ($\sim$~120 \ls/\ms) for the most luminous galaxies (see Table 2).

Combining equation (1) with equation (9), the SFR in terms of 
the HCN luminosity is
\begin{equation}
\dot M_{\rm SFR} \approx 
1.8\times 10^{-7} (L_{\rm HCN}/\ll) \ \ms  yr^{-1}. 
\end{equation}
In terms of the dense gas mass, the star formation rate becomes
\begin{equation}
\dot M_{\rm SFR} \approx
1.8({{M_{\rm dense}}\over{10^8\ms}})({{10}\over{\alpha_{\rm HCN}}})\ \ms  yr^{-1}. 
\end{equation}
Since this is a linear relation, the HCN emission is a direct  
tracer of the SFR in all galaxies. The dense gas characteristic
depletion time (half-life) is
\begin{equation}
\tau_{1/2}=0.5 M_{\rm dense}/\dot M_{\rm SFR}\approx 2.7\alpha_{\rm HCN} Myrs.
\end{equation}
Although we adopted $\alpha_{\rm HCN}\sim 10 \ms/\ll$ for normal
spirals (Paper I), this conversion
factor might be smaller for ULIGs as the $T_b$ can be
quite high (Downes \& Solomon 1998), leading to shorter dense 
gas depletion time-scales.
HCN observations could potentially become one of the best SFR tracers
in galaxies in the nearby and distant universe given the high sensitivity 
and the high spatial resolution achievable at millimeter wavelengths 
with the next generation of the millimeter telescopes.

There also appears to be some correlation between HCN emission and 
HCO$^+$, CS, and other tracers of star formation, \eg, [C~II] line emission 
and the FIR and 20cm continuum emission (Nguyen-Q-Rieu \etal 1992), 
although their sample is very limited. However, it has become clear
that [C~II] is underluminous in ULIGs (\eg, Luhman \etal 2003)
and is not a consistent star formation indicator. In addition, 
in equation (9), it was assumed that most of the \lir~ originates from
star formation with little contribution from active galactic nuclei (AGNs) 
and/or from the
general interstellar radiation field. It will be interesting 
to examine correlations between HCN and other indicators of
star formation in a 
large sample of galaxies to assess which best indicates the SFR.

The tight correlation between the IR and HCN emission also implies that 
the dominant IR emission originates from the HCN emission region, 
especially in LIGs/ULIGs with concentrated molecular gas distribution.
We know little of the size scales of the FIR emission regions 
in LIGs/ULIGs. The dominant contribution of the
radio continuum and mid-IR emission in most LIGs appears to
be from the inner regions (Condon \etal 1991; Telesco, 
Dressel, \& Wolstencroft 1993; Hwang \etal 1999; Xu \etal 2000;
Soifer \etal 2001). CO emission is usually 
concentrated in the inner regions, typically 
within a ~kpc~ of the center for ULIGs/LIGs 
(\eg, Scoville, Yun, \& Bryant 1997; Downes \& Solomon 1998; 
Sakamoto \etal 1999; Bryant \& Scoville 1999; cf. Gao \etal 1999).
HCN emission originates from the dense cores of the CO 
emitting regions, the sites of star formation, and the source
of the FIR emission. Thus, we may predict the size scales 
and location of the FIR emission by determining the HCN 
source sizes from the HCN mapping. 

\subsection{The Global Star Formation Law}

The IR--HCN linear correlation is valid  over 3 orders of magnitude from low
IR luminosity to the most luminous galaxies in the local universe. The direct
consequence of  the linear IR--HCN correlation is that the star formation law
in terms of {\it dense} molecular gas content has a power-law index of 1.0.
The global SFR  is linearly proportional to the mass of the dense molecular
gas (eq. [11] and Fig.~6). Parametrization in terms of observable mean 
surface densities of the dense
molecular gas and the SFR will not change the slope of 1 in the IR--HCN
correlation (SFR--$M_{\rm dense}$ correlation), or the linear power 
index of the star formation law, as both
quantities are simply normalized by the same galaxy disk area. Our finding of
an SFR proportional to the first power of the dense gas mass is
different from the widely used star formation law with a slope of 1.4, derived
by Kennicutt (1998a) for the disk-averaged SFR as a function of the total 
(HI and ${\rm H_2}$) or just molecular gas surface density traced 
by CO emission. As we show below, this law is not valid for normal 
spiral galaxies and results only by combining normal galaxies with 
starburst galaxies and ULIGs.

\subsubsection{Normal Spiral Galaxies}

The IR--CO correlation (SFR--$M_{\rm H_2}$ correlation) is essentially
linear up to luminosity $ L_{\rm IR} = 10^{11}$ \ls  (ULIGs and most LIGs
excluded, see Fig.~1b).  This seems to also be true in terms of the mean
surface densities of the SFR and molecular  gas mass for the nearest galaxies
with spatially resolved observations (\eg, Wong \& Blitz 2002; 
Rownd \& Young 1999). The linear IR--CO correlation at low to moderate 
IR luminosity is expected since we find that the  HCN--CO correlation 
is extremely tight and linear for normal spiral galaxies (Fig.~3).  
Thus, the linear form of the global star formation law in terms of 
total molecular gas density as traced by CO for normal galaxies is 
due to the constant dense gas mass fraction indicated by the
HCN/CO ratio (discussed in the next section).  For normal star-forming 
spirals, the star formation law is linear in terms of both the total 
molecular gas and the dense molecular gas. Then how did 
Kennicutt (1998a) obtain a slope of 1.4 in the fitting of the star 
formation law? In our  sample, a fit for the normal
galaxies in the IR--CO correlation gives a slope of 1.0 (Fig. 1b), 
but this is not the case in Kennicutt's normal galaxy sample, where 
there is a poor correlation between the SFR and the gas surface density.  
Thus, no reasonable slope can be derived from his normal galaxy sample alone.

\subsubsection{All Galaxies: Normal Spiral, Starburst, and 
Ultraluminous Galaxies}

A direct orthogonal regression fit of the IR--CO correlation for all galaxies
in our sample (Fig.~1b) leads to a slope of 1.44. But the best-fitting
least-squares slope (errors in \lir \ only) is 1.27 (1.25 if galaxies with HCN
limits are excluded, see \S 3.1). These fit slopes of the IR--CO correlation
are almost identical to the star formation power-law index in Kennicutt's
(1998a) 36 circumnuclear starbursts and ULIG sample.  It is obvious from 
Figure~1b that only galaxies with \lir $>10^{11}$ \ls (ULIGs and most LIGs) 
lie above the fixed line of slope 1.  This combination of normal and 
very luminous galaxies leads to a fit with a power-law index of 1.4. 
Therefore, this slope is not a universal slope at all as it changes 
according to the sample selection. The 1.4 slope of the composite 
Schmidt law (Kennicutt's  Fig.~6, 1998a) is  determined almost entirely 
from the starburst sample. The circumnuclear starbursts have some of 
the characteristics of ULIGs/LIGs. In particular, they must have a high 
dense gas fraction  indicated by HCN/CO ratios (see section \S 4.4).

Indeed, when we add more LIGs/ULIGs into the sample for the 
IR--CO correlation, the
slope becomes steeper. In Figure~7, we present an SFR--$M(H_2)$  
(IR--CO) correlation diagram with an additional 40 galaxies, 
mostly ULIGs, with CO data  from the literature
(Solomon et al. 1997; Gao \& Solomon 1999; Mirabel et al. 1990; 
Sanders et al. 1991), in addition to our HCN sample shown in 
Figure 1b. The least-squares fit gives a steep slope
of 1.53, and the orthogonal regression fit leads to a much 
steeper slope of 1.73.
In this sample luminous galaxies and normal galaxies have equal weight. It is
clear that the ULIGs steepen the slope of the sample.  There also appears to be
a trend that some normal spirals with the lowest $\Sigma_{\rm SFR}$ and
$\Sigma_{\rm H_2}$ in Kennicutt's sample tend to lie below the 1.4 power fit
line. Adding more extreme galaxies, both luminous ULIGs and  low-luminosity
spirals, tends to steepen the  slope  further toward 2. Therefore, it is
difficult to derive a unique 1.4 power law based upon the total molecular gas
or the total gas content.

The star formation rate in a galaxy depends linearly on the dense 
molecular gas
content as traced by HCN, regardless of the galaxy luminosity or the presence
of a ``starburst'', and not the total molecular gas and/or atomic gas
traced by CO and/or HI observations, respectively. Since dense molecular 
cloud cores are the  sites of high-mass star formation, it is the 
physical properties, location, and mass of these cores
that set the star formation rate. A detailed star formation law can be
determined from observations directly probing the Milky Way cloud cores,
particularly in the Milky Way molecular ring with spatially resolved
measurements. The molecular tracers that best quantitatively indicate the
presence of a starburst are  primarily abundant molecules with high dipole
moments such as HCN and CS requiring high molecular hydrogen density for
excitation. The molecular property that best characterizes the star formation
rate  of a galaxy is the mass of dense gas. The gas density traced by HCN
emission is apparently at or near the threshold for rapid star formation.

\subsection{HCN/CO Ratio --- A Better Indicator of Starbursts}

Although HCN luminosity is a better indicator of 
star formation than CO in galaxies, it is very useful 
to compare the HCN with CO to obtain the HCN/CO ratio.
The HCN/CO ratio is an indicator of the fraction 
of dense molecular gas available for vigorous star formation
and gauges the globally averaged molecular gas density.
The HCN/CO ratio is also a very successful predictor
of starbursts (Fig.~4) and directly correlates with the SFE 
(\lir/\lco) (Fig.~5a).  The SFE increases dramatically from {\it dense}
molecular  gas-poor galaxies to {\it dense} molecular gas-rich 
galaxies.  The global luminosity ratio 
$L_{\rm HCN}/L_{\rm CO} = <$SBR$>$ differs 
among galaxies of different luminosities (see Tables~1
\& 2) with an average ratio of 1/6 for ULIGs and only 1/25 for normal 
spirals. {\it All galaxies in our sample with a global dense gas 
mass fraction $L_{\rm HCN}/L_{\rm CO} > 0.06$ are LIGs or ULIGs} 
(Fig.~4). We note that 
Curran \etal (2000) also find an average global ratio
of 1/6  in luminous IR Seyfert galaxies, which they
attribute to star formation. 

The HCN/CO surface brightness ratio (SBR) is potentially a better 
and more practical indicator than the IR/CO ratio (the standard 
SFE diagnostic) of the starburst strength. Using the ratio  IR/CO
as a diagnostic, the IR emission is presumed to entirely originate
from star formation.  Other possible sources  such as 
AGNs and the general interstellar radiation field are assumed to 
be negligible. The HCN/CO ratio instead is directly related 
to the molecular gas properties, 
particularly the local molecular gas density, 
which is tied to star formation.  Moreover, the projection and 
confusion  of different velocities along the line of sight 
is inevitably present in IR maps, whereas the different velocity 
components  can be distinguished from the kinematics obtained
in the detailed CO and HCN maps. In addition, the low spatial 
resolution available in the FIR, even with the {\it Spitzer Space 
Telescope} and the upcoming
{\it SOFIA}, and other future far-IR space missions, is 
incompatiable with the high resolution available from millimeter 
interferometers. Therefore, it is important 
to map the HCN/CO ratio in galaxies in order to fully explore 
the star formation properties and SFE.

As we show elsewhere from  HCN maps of
nearby galaxies (Gao 1996; Y. Gao \& P.M. Solomon 2004, in preparation),
 the ratio of $I_{\rm HCN}/I_{\rm CO}$, \ie, 
the surface brightness ratio (SBR),
in most cases, is high in the centers of
spiral galaxies and typically drops off at large galactic radii. 
A significant fraction of dense molecular gas is (traced by 
HCN emission) distributed in the inner disks of galaxies
outside  the nuclear or inner ring starburst regions and can
be detected to radii as large as a few kpc, perhaps to diameters
of $\sim D_{\rm 25}/4$. We find the highest fraction of dense molecular 
gas, indicated by the SBR (as measured by
single-dish telescopes with beam sizes $\sim 1$ kpc),
\begin{equation}
{\rm SBR}\ ({\rm cores}) \equiv I_{\rm HCN}/I_{\rm CO} \ ({\rm cores}) 
\approx  0.1,
\end{equation}
nearly comparable to those observed globally from ULIGs, in the 
centers of most normal spiral galaxies observed (usually the 
nuclear starburst cores). 
We attribute this to the presence of a starburst. 
Helfer \& Blitz (1993, 1997a) found similarly high ratios in 
the centers of normal galaxies, which they relate to the high 
ambient pressure, but they do not correlate their data with
infrared luminosity or star formation rates. 

We also find that the SBR ratio 
generally falls off in the disks at larger radii (\eg,
$\approxgt 3$ kpc), to a very low SBR
$\sim$ 0.015--0.03. This low SBR (disks)~$\approxlt 0.03$ is
the same as that found in the Milky Way's disk  on average, and 
over the full extent of nearby GMCs in the Milky Way 
(Helfer \& Blitz 1997b), as well as outside the central regions 
in several normal spiral galaxies mapped with an interferometer 
(Helfer \& Blitz 1997a). 

Although the {\it global} luminosity ratio 
$L_{\rm HCN}/L_{\rm CO} = <$SBR$>$ differs 
dramatically among galaxies of different luminosities 
(see Tables~1 \& 2), the difference between the SBR in central 
beam measurements of galaxies is  related to 
the different telescope beam diameters (\eg, Helfer \& Blitz 1993),
different source sizes (of CO and HCN), and different
size scales of the starburst cores in the centers of individual 
galaxies. The CO/HCN intensity 
ratio (the inverse of SBR) changes from ~20 -- 80
for normal spiral galaxies to ~4 -- 10 for ULIGs.  
However, if only central beam measurements are used, the ratio seems 
quite uniform (Aalto \etal 1995). 
The central beam measurement is not
representative of the global measurment (of the entire
galaxy) and may not even be an accurate measurement of the 
central region since it is telescope dependent. This explains why
Aalto \etal (1995) did not find the strong $L_{\rm IR}$--\lhcn~ correlation. 

The distribution of the HCN emission, particularly HCN/CO
ratio maps, can be used directly to 
locate starburst sites in nearby galaxies. High-resolution maps
of HCN and HCN/CO have been obtained in the central regions of a 
few nearby normal galaxies, some with clear central starbursts.
Paglione \etal (1995) mapped the central bar in the starburst 
nucleus of NGC~253 and found a peak HCN/CO=0.2 falling off to 
0.04 at a galactocentric radius of 200 pc. The peak SBR in this starburst
is similar to the global value in the most luminous galaxies in our
sample (Fig.~4). Both the strongest HCN surface brightness and
the strongest SBR occurred at the location of strong, extended
nonthermal radio continuum and thermal free-free emission
associated with the starburst.
Using our Figure~5a, this high SBR (0.15--0.2) indicates that this small
central starburst has an \lir/\lco~ ratio comparable to that of ULIGs,
\lir/\lco$\sim 200 \ls/\ll$. Unfortunately, there are no high-resolution
IR observations available to test this prediction. Paglione \etal 
(1995) estimated the \lir~ from the radio continuum to be 
\lir$\sim 1\times 10^9 \ls$. The \lco~ estimated from the HCN
and HCN/CO ratios is approximately $4\times 10^7\ll$ yielding
a rough $\lir/\lco \sim 250\ls/\ll$, in agreement with the prediction
of Figure~5a.

High HCN/CO ratios (0.1--0.2) are also found in the central 
$\approxlt 1$~kpc regions of several other galaxies (Downes \etal 1992; 
Helfer \& Blitz 1997a; Reynaud \& Downes 1997; 
Kohno \etal 1996; Kohno, Kawabe, \& Vila-Vilaro 1999). 
Kohno \etal (1999) found that most
of the HCN emission, as well as an enhanced HCN/CO ratio, 
is associated with the circumnuclear star-forming ring in NGC~6951.
No significant enhancement of the HCN/CO ratio is observed at
the CO peaks. In IC~342 with a resolution of 60 pc\footnote{Distance
to IC 342 is adopted as 3.7 Mpc, see Table 1.}  
(Downes \etal 1992), however, only three of five HCN clouds
seem to be actively forming into stars based on the
presence of free-free emission. However, the observations of the
free-free emission are still of limited sensitivities, either
at 2cm and 6cm (Turner \& Ho 1983; Turner \etal 1993) or at 3 mm 
(Meier \& Turner 2001). And there are no high-resolution FIR observations 
available to clearly indicate the star formation activities and 
star formation rates at this small scale. Judging from all these
observations and the $H\alpha$ emission (Turner \& Hurt 1992), there 
is probably star formation in all the HCN clouds (D. Downes 2003, 
private communication).
Are some of these HCN clouds precursors to a starburst? 
Although the HCN/CO ratios are high ($\sim 0.14$) for these 
HCN clouds, the average of a several hundred parsec central 
region has only an HCN/CO ratio of $\approxlt 0.05$ in IC~342,
clearly not a strong starburst.

A recent summary of the HCN/CO ratios obtained in six central
nuclear starbursts (Shibatsuka \etal 2003) lists a typical
ratio HCN/CO$\approx$0.1--0.25. Sorai \etal (2002) also
find the highest HCN/CO ratio of $\sim 0.1$ in central
regions of a few nearby galaxies. This is the same as the global
ratio we find in many LIGs and all ULIGs that
have HCN luminosities several hundred times greater than
the central starbursts of normal galaxies.

The only LIGs that have been well imaged in 
HCN are the AGN/starburst hybrid galaxy NGC~1068 
(Tacconi \etal 1997, 1994; Helfer \& Blitz 1995; Jackson \etal 1993), 
which has very high HCN/CO ratio ($\approxgt 0.3$) within $\sim 100$~pc
nuclear region, the
merging pair Arp~299 (Aalto et al. 1997; Casoli et al. 1999), 
and the archetypal ULIG Arp~220 (Radford \etal 1991; 
N.~Z. Scoville 2001, private communication). Arp~299 may 
harbor an AGN in the eastern nucleus (\eg, Gehrz et al. 1983),
similar to NGC~1068. It appears that some weak Seyferts also 
have rather high HCN/CO ratio ($\approxgt 0.2$) in the innermost 
$\approxlt $100~pc nuclear region around the AGN (\eg, M51, Kohno \etal 1996). 
But this contributes little to the average HCN/CO ratios
in the circumnuclear ($\sim $0.5--1~kpc) starburst rings where 
most of the molecular gas is located.

Ultimately, it is the dense molecular 
gas, rather than the total molecular gas content,
that is the raw material for active star formation in galaxies.
Our global measurements of 65 galaxies show that the dense molecular
gas fraction indicated by \lhcn/\lco~ is an important measure
of the star formation efficiency; the mass of dense molecular
gas indicated by ~\lhcn~ is a very good measure of the star
formation rate deduced from \lir. The total molecular content
indicated by \lco~ is an unreliable indicator of star formation
rate particularly in starburst galaxies.
The location of starburst regions and better characterization 
of the starbursts in individual galaxies could be 
better indicated by the local SBR measurements, obtained by
the high-resolution and high-sensitivity HCN and CO observations.

\subsection{Starburst Origin of the Far-IR Emission}

Our results show that high-mass star formation in dense molecular
gas is responsible for the infrared luminosity from a wide range
of galaxies including normal spirals of moderate IR luminosity
($5\times 10^9 \ls \approxlt \lir \approxlt 10^{11} \ls$),
luminous infrared galaxies (LIGs, $10^{11} \ls \approxlt \lir 
\approxlt 10^{12} \ls$), and ultraluminous galaxies (ULIGs,
$\lir \approxgt 10^{12} \ls$). The star formation rate or
IR luminosity is proportional to the mass of dense molecular
gas in all galaxies.

\subsubsection{Luminous and Ultraluminous Infrared Galaxies}

Evidence is mounting that the dominant energy
source in most ULIGs is from the extreme starbursts rather than
the dust-enshrouded AGNs (\eg, Solomon \etal 1997;
Downes \& Solomon 1998; Genzel et al. 1998; 
Scoville et al. 2000; Soifer \etal 2001).
The debate about the energy source in ULIGs, particularly 
in ``warm'' ULIGs, has been going on for over a decade
(\eg, Sanders \etal 1988; Veilleux et al. 1999; Sanders 1999; 
Joseph 1999). Although AGNs are present in many ULIGs (\eg,
Nagar \etal 2003; Franceschini \etal 2003), there is
little evidence indicating that the 
AGN contribution is the dominant source of the FIR emission.
 
Our results summarized in \S~3 provide compelling evidence 
in favor of a star formation origin for the huge infrared luminosity 
from ultraluminous galaxies. We have shown that the infrared 
luminosity of all molecular gas-rich
spiral galaxies including ULIGs is proportional to the dense star-forming
molecular gas mass traced by \lhcn (see Fig. 1a and Fig. 6).
 
One of the main arguments in favor of an AGN as the power source in 
ULIGs is the anomalously high ratio \lir/\lco~ or \lir/M(H$_2$).  
The IR luminosity and thus
the required star formation rate per solar mass of molecular gas 
(as traced by CO emission) is an order of magnitude higher in 
ultraluminous and most luminous galaxies than in normal spiral 
galaxies, suggesting that an AGN rather than star formation 
is required (Sanders \etal 1988, 1991)
in order to produce the high infrared emission.  Viewed in
terms of the dense gas mass the situation is completely different. 
The ratio \lir/\lhcn~ or \lir/M$_{\rm dense}$  is essentially 
the same in all galaxies including ULIGs. Figure 8
shows that \lir/M$_{\rm dense}$  is virtually independent of 
galaxy luminosity and on average \lir/M$_{\rm dense} = 90 \ls/\ms$, 
about the same as in molecular cloud cores
but much higher than in GMCs as a whole (Mooney \& Solomon 1988).
Ultraluminous galaxies simply have a large quantity of dense molecular 
gas and thus produce a prodigious starburst that heats the dust. 
It is not surprising that starbursts of
this magnitude are observed in the infrared and never seen in the optical-UV
part of the spectrum. The young OB stars are imbedded in very massive 
and dense regions dwarfing anything found in Milky Way GMCs and all 
of their optical-UV radiation is absorbed by dust. Although ULIGs 
are not simply scaled-up versions
of normal spirals in terms of their total molecular mass, they are scaled-up
versions in terms of their dense molecular gas mass, which is exactly what is
expected if star formation is the power source. 
                                              
Ultimately, even for warm ULIGs (\eg, Surace \etal 1998), 
which might have, to some degree, evolved into the phase of the
dust-enshrouded QSOs/AGNs (\eg, Sanders \etal 1989; 
Surace \& Sanders 1999; Veilleux \etal 1999; Genzel \etal 1998; 
Sanders 1999), we may still be able to tell whether starbursts still 
dominate most of the high $L_{\rm IR}$ or not by
examining their total dense molecular gas content and the fraction of 
dense molecular gas. For instance, \IRAS~05189-2524 and Mrk 231 are
warm ULIGs, but they have similar \lir/M$_{\rm dense}$ and HCN/CO
ratio as the other seven out of nine ULIGs in our sample. Clearly, 
more HCN observations are required 
to judge if warm ULIGs distinguish themselves from other ULIGs.

\subsubsection{Normal Spiral Galaxies}

Although some fraction of \lir~ originates
from outer disks in nearby large spiral galaxies, the 
dominant contribution is still from the centers 
(\eg, Rice \etal 1988). Better resolution \IRAS maps, obtained by
an improved imaging deconvolution algorithm, tend to show much more 
centrally concentrated FIR emission in the inner disks 
(Rice 1993). Devereux \& Young (1990) have shown that the global 
FIR luminosity in spiral galaxies is consistent with that 
contributed by warm dust heating from the high-mass OB stars.
Higher resolution (as compared with that of \IRAS) 
measurements of the 160 $\mu$m and H$\alpha$ 
emission in nearby galaxies NGC~6946 and M51 suggest that the 
FIR luminosity is in quantitative agreement with that expected 
from OB stars throughout 
the star-forming disks (Devereux \& Young 1992, 1993). 

The existence of cold dust components of $\la 30$~ and $\la 20$~K 
in nearby galaxies has been revealed from both recent 
ISOPHOT FIR observations (\eg, Popescu \etal 2004; 
Haas \etal 1998; Alton \etal 1998a; Davies \etal 1999) 
and SCUBA submillimeter observations (\eg, Alton \etal 1998b; 2000),
respectively. The cold dust distribution usually has a larger radial 
extent (\eg, Alton \etal 1998a) and the dominant cold dust component 
may not be closely associated
with the active star-forming inner disks. HCN maps of nearby galaxies 
(\eg, Nguyen-Q-Rieu \etal 1992; Helfer \& Blitz 1997a; Y. Gao \& 
P.M. Solomon 2004, in preparation) show that the HCN emission region is 
much more compact than that of CO emission region in normal spiral
galaxies and/or starburst galaxies and is thus closely related to
the warm dust in the inner disks. 

The contribution of the general interstellar radiation 
field to the total FIR emission
in galaxies might be significant at the low-\lir~ end, where 
the general infrared interstellar radiation field is comparable to 
the IR radiation from the active star formation. 
This might be testable on the IR--HCN correlation plot (Fig.~1a)
for galaxies of the lowest \lir~ ($\la 5\times 10^9 \ls$) when
more observations of the lowest \lir~ sources are available to 
show a statistically significant trend. 
Given that the tight correlation between these two quantities 
might be used to predict one another, one can check whether 
low-\lir~ galaxies have a higher \lir/\lhcn~ than expected
from the IR--HCN correlation.

\subsection{High Redshift Galaxies and AGN}

AGNs may become more important than star formation at the very 
high IR luminosity end,
especially in warm ULIGs (Veilluex \etal 2003),  
in contributing the bulk of the energy output 
(\eg, Sanders \etal 1988; Veilluex \etal 1999). For extremely luminous
galaxies with \lir$\sim 10^{13}\ls$, the so-called hyperluminous
infrared galaxies (HLIGs, Sanders \& Mirabel 1996; Rowan-Robinson 2000),
the implied SFR (Equation 9) would be $\sim 2000~\ms yr^{-1}$.
Although such super-starburst
galaxies likely exist at high redshifts and HLIGs 
could indeed be such super-starburst systems (if the
magnification by a gravitational lens and the AGN contribution
to the IR emission are not important), there are no similar 
systems in the local universe. 

If the tight correlations obtained
from our local HCN sample are applicable to high-z galaxies, 
we can roughly estimate their expected molecular gas properties.
For the whole sample \lir/\lhcn=900\ls/\ll~ (Equation 1). 
But ULIGs have a slightly higher ratio of
\lir/\lhcn=1200\ls/\ll~ (Table 2). 
For high-z galaxies with \lir$\sim 10^{13}\ls$, 
we thus expect to have $\lhcn \sim 0.8 \times 10^{10}\ll$ if
they are the analogs of local ULIGs.
This is only a factor of 2--3 higher than the highest
\lhcn~ observed in local ULIGs.
The first high-z HCN detection is from the 
Cloverleaf quasar at z=2.567, helped by the magnification 
of the gravitational lens, with the intrinsic, 
magnification-corrected $\lhcn \sim 0.3 \times 10^{10}\ll$ 
(Solomon \etal 2003). It appears that, if the hot dust AGN component
can be subtracted from the total IR emission, then even the 
Cloverleaf quasar seems to fit the IR--HCN correlation determined
here from the local universe. 

At present, more than 20 CO sources at high $z$ 
have been detected, all with high CO luminosity
(\eg, Cox \etal 2002; Carilli \etal 2002; 
Papadopoulos \etal 2001; Guilloteau \etal 1999; Downes \etal 1999; 
Frayer \etal 1998). 
We can estimate the expected \lhcn~ for these extraordinarily 
large \lco~ sources detected at high $z$ using our HCN--CO 
correlation (Fig.~3). Cox \etal (2002) report 
the strongest CO emitter detected to date, with $\lco\sim 10^{11}\ll$ 
even after the magnification by the gravitational lensing 
is corrected. Using Equation 2, or
$\lhcn/\lco=0.1\times (\lco/10^{10}\ll)^{0.38}$, we obtain 
$\lhcn \sim 2.4\times 10^{10}\ll$ and \lhcn/\lco=0.24. 
This \lhcn/\lco~ ratio is the same as the very highest found for local
ULIGs. For this largest \lco~ source, \lir$\sim 2.0\times 10^{13}\ls$ 
is estimated 
by Cox \etal (2002) based on the 250 and 350 GHz measurements 
(Omont \etal 2001; Isaak \etal 2002). And using \lir/\lhcn=1200\ls/\ll,
we expect to have $\lhcn=1.7\times 10^{10}$\ll. Perhaps a
better estimate of the expected HCN luminosity can be constrained 
from both IR and CO luminosities by using the multiparameter fit 
from equation (4). Indeed, we also obtain
$\lhcn=1.7\times 10^{10}\ll$. If such high ~\lhcn~ is eventually
detected in this source, then star formation rather than AGNs must
be responsible for most of the high infrared luminosity.

We can also roughly estimate the expected upper limit of \lir/\lco~
for HLIGs and/or high-z galaxies, if they are powered by star 
formation and the correlation 
between \lir/\lco~ and \lhcn/\lco~ found in Figure~5a is applicable. 
Although there are very large scatters ($\sim 0.5$ dex, $2\sigma$) 
in the fit, 
this should be good to within a factor of $\sim 3$. Here we can take 
$L_{\rm HCN}/L_{\rm CO}\approxlt 1$ as upper limits. 
Using the orthogonal fit $log\lir/\lco=1.24log\lhcn/\lco + 3.24$, 
we obtain \lir/\lco$\approxlt 1700 \ls/\ll$. Therefore, the 
expected maximum \lir~ from star formation
is always less than $\approxlt 1.7\times 10^{14} (\lco/10^{11}\ll) \ls$. 
 
Although many submillimeter galaxies
are HLIGs, few have $\lir\sim 10^{14}\ls$ (Chapman \etal 2003).
The highest \lco~ detected among submillimeter galaxies is
$\sim 0.7\times 10^{11}\ll $ (Neri \etal 2003; Greve, 
Ivison, \& Papadopoulos 2003). If any sources indeed have high 
\lhcn$\sim 0.7\times 10^{11}\ll$, 
it appears that extreme starbursts from abundant active star-forming 
dense molecular gas are still possible to power such 
$\lir\sim 10^{14}\ls$ sources. This corresponds roughly to the 
maximum possible SFR of $\sim 2\times 10^4$\ms/yr (using eq. [9]), 
as argued by Heckman (2000) based on 
physical causality for self-gravitating, extremely massive 
galaxy systems of $\sim 10 \times L_*$. Observing HCN in addition 
to CO will help distinguish whether star formation rather than AGNs 
contributes mostly to the highest \lir~ observed.

\section{SUMMARY}

HCN luminosity is a tracer of {\it dense} molecular gas, 
$n(H_2) \approxgt 3 \times 10^4$cm$^{-3}$, associated with 
star-forming molecular 
cloud cores. Here we briefly summarize the principal results found from
an analysis of our HCN survey of galaxies:

1. A tight linear correlation between the IR and HCN luminosities
$L_{\rm IR}$ and $L_{\rm HCN}$ in 65 galaxies is established
with a correlation coefficient R=0.94 (Fig. 1a). There is also
a significant correlation between the normalized luminosities
$L_{\rm IR}/L_{\rm CO}$ and $L_{\rm HCN}/L_{\rm CO}$ which confirms
the true physical relationship between  \lir~ and \lhcn.          
The IR--HCN linear correlation is valid  over 3 orders of magnitude 
including ultraluminous infrared galaxies, the most 
luminous galaxies in the local universe. The direct
consequence of  the linear IR--HCN correlation is that the star formation law
in terms of {\it dense} molecular gas content has a power law index of 1.0.
The global star formation rate  is linearly proportional to the mass 
of  dense molecular gas in normal spiral galaxies, luminous infrared
galaxies and ultraluminous infrared galaxies. This is strong evidence 
in favor of star formation as the power source in 
ultraluminous infrared galaxies since the star formation 
in these  galaxies appears to be normal and expected given their 
high mass of dense star-forming gas. 

2. The star formation rate indicated by \lir~ depends on the amount of 
dense molecular gas traced by HCN emission, not the total molecular 
gas traced by CO emission. One of the main arguments in favor of an 
AGN (Sanders \etal 1988) as the power source in ultraluminous 
infrared galaxies is the anomalously high ratio \lir/\lco~ or \lir/M(H$_2$. 
The IR luminosity and thus
the required star formation rate per solar mass of molecular gas 
traced by CO emission, is an order of magnitude higher in
ultraluminous infrared galaxies and most luminous infrared 
galaxies than in normal spiral galaxies. This has been interpreted 
as indicating that a dust enshrouded AGN
rather than star formation is required to produce the very high luminosity
(Sanders \etal 1991).  Viewed in
terms of the dense gas mass the situation is completely different.
The ratio \lir/\lhcn~ or \lir/M$_{\rm dense}$  is the same in all
galaxies including ultraluminous infrared galaxies. Figure 8
shows that \lir/M$_{\rm dense}$  is virtually independent of
galaxy luminosity and on average \lir/M$_{\rm dense} \approx 90 \ls/\ms$,
about the same as in molecular cloud cores
but much higher than in GMCs as a whole (Mooney \& Solomon 1988). 
Ultraluminous infrared galaxies simply have a large quantity of 
dense molecular gas and thus produce a prodigious starburst which 
heats the dust, produces the IR, and blocks all or most optical radiation. 
Although ultraluminous infrared galaxies are not simply scaled up versions
of normal spirals in terms of their total molecular mass they are scaled up
version in terms of their dense molecular gas mass, which is exactly what 
is expected if star formation is the power source. We note that our 
sample includes 9 ultraluminous infrared galaxies 
(\lir$> 0.8\times 10^{12} \ls$) and 23 luminous infrared galaxies 
(\lir$>10^{11}\ls$) but only 2 ``warm'' ultraluminous infrared 
galaxies (Surace \etal 1998), a subclass representing about 25\% 
of all ultraluminous infrared galaxies. Although these 2 warm 
ultraluminous infrared galaxies 
appear normal in terms of the IR/HCN ratio, further investigation 
of warm ultraluminous infrared galaxies is required.

3.  The HCN/CO ratio is an indicator of the fraction
of dense molecular gas available for vigorous star formation
and gauges the globally averaged molecular gas density. 
It is a powerful starburst indicator.

There is a strong correlation between the HCN and CO luminosities 
in galaxies although the correlation is not linear.  
The ratio $L_{\rm HCN}/L_{\rm CO}$ is
constant for normal spirals and increases for luminous and ultraluminous 
IR galaxies. The global ratio  $L_{\rm HCN}/L_{\rm CO} = <$SBR$>$  
(see Tables~1 \& 2) has an average ratio of 1/6 for ultraluminous 
infrared galaxies and only 1/25 for normal spirals.
The HCN/CO ratio is  a very successful predictor
of starbursts (Fig.~4) and directly correlates with the star formation
efficiency indicator
(\lir/\lco) (Fig.~5a).  The SFE increases dramatically from {\it dense}
molecular  gas-poor galaxies to {\it dense} molecular gas-rich
galaxies. {\it All galaxies in our sample with a high dense gas 
mass fraction indicated by $L_{\rm HCN}/L_{\rm CO} > 0.06$ are 
luminous or ultraluminous infrared galaxies} (Fig.~4).
                                                                       
4. The correlation between $L_{\rm IR}$ and $L_{\rm CO}$
may  be a consequence of the  stronger  correlations between 
$L_{\rm IR}$ and $L_{\rm HCN}$ and between $L_{\rm HCN}$ and 
$L_{\rm CO}$.  A model two parameter fit for
\lir(\lhcn, \lco) shows almost no dependence on \lco. Much of 
the CO emission originates from moderate density regions with 
low to moderate dust temperature and little or no active high 
mass star formation. A critical molecular parameter that 
measures star formation rates in galaxies is the amount of
dense molecular gas measured by the HCN luminosity.
High molecular gas density, strong HCN emission and 
warm dust heated by the newly formed OB stars go together.  
                                            
5. A quantitative star formation theory must start with the 
processes that form dense cloud cores, particularly massive 
cores. It appears from our survey of 65 IR/CO bright galaxies 
including normal spirals and very luminous IR starbursts that once
the local density is raised from that of a typical GMC ($n(H_2) \sim$ 
a few $\times 10^2$~cm$^{-3}$) to $3 \times 10^4$cm$^{-3}$, 
star formation including high-mass star formation
is efficient and progresses rapidly. The characteristic time 
scale for using half of all the dense gas is about $2\times 10^7$ years.
                                                                               
\acknowledgments

Many thanks are owed to Judy Young for
helpful discussions and suggestions.
We appreciate the generous support and allocation of observing time
from the NRAO 12m, IRAM 30m, and FCRAO 14m telescopes. We also
thank the anonymous referee for a careful and helpful report. 
This research has
made use of the NASA/IPAC Extragalactic Database (NED), which is
operated by the
Jet Propulsion Laboratory, Caltech, under contract with the National
Aeronautics and Space Administration.

\clearpage

\begin{deluxetable}{lrrrrrrr}
\tablenum{1}
\tablecolumns{6}
\tablecaption{Global Properties of Galaxies in the HCN 
Survey\tablenotemark{a} \label{tbl-1}}
\tablehead{
\colhead{Galaxies}            &       \colhead{$D_L$}          &
\colhead{$L_{\rm IR}$}        &       \colhead{$L_{\rm CO}$}   &
\colhead{$L_{\rm HCN}$\tablenotemark{b}}    &   
\colhead{$L_{\rm HCN}/L_{\rm CO}$}          &
\colhead{\lir/\lhcn} & \colhead{$T_{\rm dust}$\tablenotemark{c}} \\
\colhead{   }                 &       \colhead{Mpc}            &
\colhead{$10^{10}$ \ls}       &  
\multicolumn{2}{c}{$10^8~{\rm K \kms pc}^2$}   &   
\colhead{$<$SBR$>$}           &
\colhead{\ls/\ll}             &        \colhead{K}}
 
\startdata

NGC 253      &   2.5 &  2.1 &    4.6 &    0.27 &   0.059 & 778 &34\nl
{\bf IC 1623}&  81.7 & 46.7 &  130.5 &     8.5 &   0.065 & 549 &39\nl
NGC 660      &  14.0 &  3.7 &    7.3 &$>$0.26&$>$0.036&$<$1420 &37\nl 
{\bf NGC 695}& 133.5 & 46.6 &   92.9 &     4.3 &   0.046 &1080 &34\nl
{\bf MRK 1027}&123.5 & 25.7 &   41.7 &    1.89 &   0.045 &1350 &37\nl
NGC 891      &  10.3 &  2.6 &   11.0 &    0.25 &   0.024 &1120 &28\nl
NGC 1022     &  21.1 &  2.6 &    4.2 &    0.20 &   0.047 &1300 &39\nl
NGC 1055     &  14.8 &  2.1 &   13.3 &$<$0.37&$<$0.028& $>$568 &29\nl
{\bf NGC 1068}& 16.7 & 28.3 &   20.7 &    3.61 &   0.174 & 784 &40\nl
{\bf NGC 1144}&117.3 & 25.1 &  108.9 &    2.67 &   0.025 & 940 &32\nl
{\bf NGC 1365}& 20.8 & 12.9 &   58.7 &    3.10 &   0.053 & 420 &32\nl
IC 342       &   3.7 &  1.4 &    9.5 &    0.47 &   0.050 & 300 &30\nl
{\bf NGC 1614}& 63.2 & 38.6 &   24.5 &    1.25 &   0.051 &3090 &46\nl
{\bf *VIIZw31}&223.4 & 87.1 &  125.0 &     9.8 &   0.078 & 890 &34\nl
{\bf *05189-2524}&170.3&118.1&  67.0 &     6.2 &   0.093 &1900 &48\nl
NGC 2146     &  15.2 & 10.0 &   12.5 &    0.96 &   0.071 &1040 &38\nl
NGC 2276     &  35.5 &  6.2 &   10.2 &    0.40 &   0.039 &1550 &31\nl 
{\bf ARP 55} & 162.7 & 45.7 &  125.0 &     3.8 &   0.030 &1200 &36\nl
NGC 2903     &   6.2 & 0.83 &    2.3 &$>$0.09&$>$0.036&$<$ 922 &29\nl
{\bf *UGC 05101}&160.2& 89.2&   50.8 &    10.0 &   0.197 & 892 &36\nl
M82          &   3.4 &  4.6 &    5.7 &    0.30 &   0.053 &1530 &45\nl
NGC 3079     &  16.2 &  4.3 &   24.0 &$\sim$1.0&$\sim$0.042&$\sim$430&32\nl
{\bf *10566+2448}&173.3& 93.8&  61.5 &    10.2 &   0.166 & 920 &42\nl
{\bf ARP 148}& 143.3 & 36.5 &$>$47.0 &    4.0 & $<$0.085 & 913 &35\nl
NGC 3556     &  10.6 & 1.35 &$>$4.5 &$>$0.09&$\sim$0.020&$<$1500&30\nl 
NGC 3627     &   7.6 & 1.26 &    4.4 & $>$0.08&$>$0.017&$<$1580&30\nl
NGC 3628     &   7.6 & 1.01 &    7.1 &    0.24 &   0.034 & 421 &30\nl
NGC 3893     &  13.9 & 1.15 &    4.1 &    0.23 &   0.056 & 500 &30\nl
NGC 4030     &  17.1 & 2.14 &   15.2 &    0.54 &   0.036 & 398 &29\nl
NGC 4041     &  18.0 & 1.70 &    3.9 &    0.18 &   0.046 & 944 &31\nl 
NGC 4414     &   9.3 & 0.81 &    4.6 &    0.16 &   0.033 & 510 &29\nl
NGC 4631     &   8.1 &  2.0 &    2.3&$\sim$0.08&$\sim$0.037&$\sim$2380&30\nl
NGC 4826     &   4.7 & 0.26 &    1.3 &$>$0.04&$>$0.030& $<$750 &33\nl
NGC 5005     &  14.0 &  1.4 &    8.2 &    0.41 &   0.049 & 350 &28\nl
NGC 5055     &   7.3 &  1.1 &    8.6 &$>$0.10&$>$0.012&$<$1140 &26\nl
{\bf NGC 5135}& 51.7 & 13.8 &   31.3 &    2.73 &   0.087 & 510 &36\nl 
M83          &   3.7 &  1.4 &    8.1 &    0.35 &   0.043 & 420 &31\nl
{\bf *MRK 273}&152.2 &129.9 &   65.0 &    15.2 &   0.234 & 860 &48\nl
\tablebreak
NGC 5678     &  27.8 &  3.0 &   17.2 &    0.75 &   0.044 & 410 &29\nl
NGC 5713     &  24.0 &  4.2 &    8.1 &    0.22 &   0.027 &1880 &27\nl
NGC 5775     &  21.3 &  3.8 &   10.9 &    0.57 &   0.052 & 670 &32\nl
{\bf *17208-0014}&173.1&234.5& 146.9 &    37.6 &   0.256 & 640 &46\nl 
{\bf 18293-3413}&72.1& 53.7 &   85.5 &    4.03 &   0.047 &1330 &38\nl 
{\bf NGC 6701}& 56.8 & 11.2 &   34.0 &    1.38 &   0.041 & 820 &32\nl
{\bf NGC 6921}& 60.3 & 11.4 &   17.5 &$\sim$2.81&$\sim$0.160&$\sim$410&34\nl
NGC 6946     &   5.5 &  1.6 &    9.2 &    0.49 &   0.053 & 330 &30\nl
{\bf NGC 7130}& 65.0 & 21.4 &   44.9 &    3.27 &   0.071 & 660 &37\nl
{\bf IC 5179}&  46.2 & 14.1 &$\sim$26.4& 3.42&$\sim$0.129& 420 &33\nl
NGC 7331     &  15.0 &  3.5 &$>$10.7&$>$0.44&$\sim$0.041&$<$800&28\nl
{\bf NGC 7469}& 67.5 & 40.7 &   37.1 &    2.19 &   0.059 &1860 &41\nl
NGC 7479      & 35.2 &  7.4 &   26.7 &    1.12 &   0.042 & 665 &36\nl
{\bf *23365+3604}&266.1&142.0&  85.0 &    15.0 &   0.176 & 950 &45\nl
{\bf MRK 331} & 75.3 & 26.9 &   52.1 &    3.35 &   0.064 & 810 &41\nl 
          &       &      &        &         &          &  &  \nl
\cutinhead{Literature HCN Data}
{\bf *MRK 231}& 170.3 &303.5&   82.2 &    18.6 &   0.226 &1630 &47\nl 
{\bf *ARP 220} &  74.7 &140.2&   78.5 &     9.2 &   0.117 &1520 &44\nl
{\bf NGC 6240}&  98.1 & 61.2&   79.0 &    11.0 &   0.139 & 560 &41\nl
{\bf ARP 193} &  92.7 & 37.3&   39.8 &     9.5 &   0.238 & 400 &37\nl
{\bf ARP 299} &  43.0 & 62.8&   29.0 &     2.1 &   0.072 &2990 &46\nl
{\bf NGC 7771}&  60.4 & 21.4&   90.8 &     6.5 &   0.070 & 335 &33\nl
{\bf NGC 828}&  75.4 & 22.4 &   58.5 &     1.3 &   0.022 &1720 &33\nl
NGC 520      &  31.1 &  8.5 &   16.3 &    0.64 &   0.039 &1330 &38\nl
NGC 3147     &  39.5 &  6.2 &   59.0 &    0.90 &   0.015 & 690 &26\nl
NGC 1530     &  35.4 &  4.7 &   23.0 &    0.49 &   0.021 & 960 &29\nl
NGC 4945     &   3.7 &  2.6 & 5.8 &$\sim$0.27&$\sim$0.047&$\sim$960 &31\nl
M51          &   9.6 &  4.2 &   19.4 &    0.50 &   0.026 & 850 &30\nl
 
\tablenotetext{a}{This table contains all HCN Survey data of Gao
\& Solomon (2003, Paper I) and includes a dozen 
galaxies in the literature (almost entirely from Solomon \etal 1992,
but M51 and NGC~4945 from Nguyen-Q-Rieu \etal 1992 and Henkel \etal
1994, respectively). LIGs with
\lir$> 10^{11}\ls$ are in boldface, and ULIGs with 
\lir$\approxgt 10^{11.9}\ls$ are further marked with an asterisk (\*).}
\tablenotetext{b}{The 2$\sigma$ 
upper limit ($<$) is listed for NGC~1055. The lower limits ($>$) are 
for nearby galaxies, where we only detected HCN in the galaxy 
centers or more extensive mapping is still required.}
\tablenotetext{c}{The warm dust temperature $T_{\rm dust}$ is for 
optical thin dust with emissivity 
$\sim \nu^\beta, \beta=1.5$.}
\enddata
\end{deluxetable}

\clearpage
    
\begin{deluxetable}{lrrrrrr}
\tablenum{2}
\tablewidth{6in}
\tablecaption{Summary of the Average IR, CO and HCN Luminosities \label{tbl-2}}
\tablehead{
\colhead{$L_{\rm IR}$}        &      \colhead{No.\tablenotemark{1} ~of} & 
\colhead{$L_{\rm HCN}$}       &       \colhead{$L_{\rm CO}$}  &
\colhead{$L_{\rm IR}/L_{\rm CO}$}          &
\colhead{$L_{\rm IR}/L_{\rm HCN}$}         &
\colhead{$L_{\rm CO}/L_{\rm HCN}$\tablenotemark{2}} \\
\colhead{\ls}                 &       \colhead{galaxies} &
\multicolumn{2}{c}{$10^8~{\rm K \kms pc}^2$}   &   
\multicolumn{2}{c}{\ls/${\rm K \kms pc}^2$}   &   
\colhead{ }}
 
\startdata

Normal$<10^{11}$           & 26 & 0.35 &  8.7 &  30 &  740 & 25\nl
$10^{11} <$LIGs$\approxlt 0.8\times 10^{12}$& 22 & 3.1  & 47.6 &  60 &  890 & 14\nl 
ULIGs\approxgt$10^{11.9}$   &  9 & 9.0  & 69.0 & 170 & 1200 &  6\nl
All                  & 57 & 1.5  & 20.0 &  50 &  870 & 17\nl
\tablenotetext{1}{Galaxies with only limits in HCN luminosity are excluded.}
\tablenotetext{2}{The inverse of average surface brightness ratio 
$<$SBR$>$ $\equiv L_{\rm HCN}/L_{\rm CO}$ indicating the fraction of 
dense molecular gas in galaxies.}
\enddata
\end{deluxetable}

\newpage

\begin{figure*}
\epsscale{0.68}
\vskip -0.5in
 \plotone{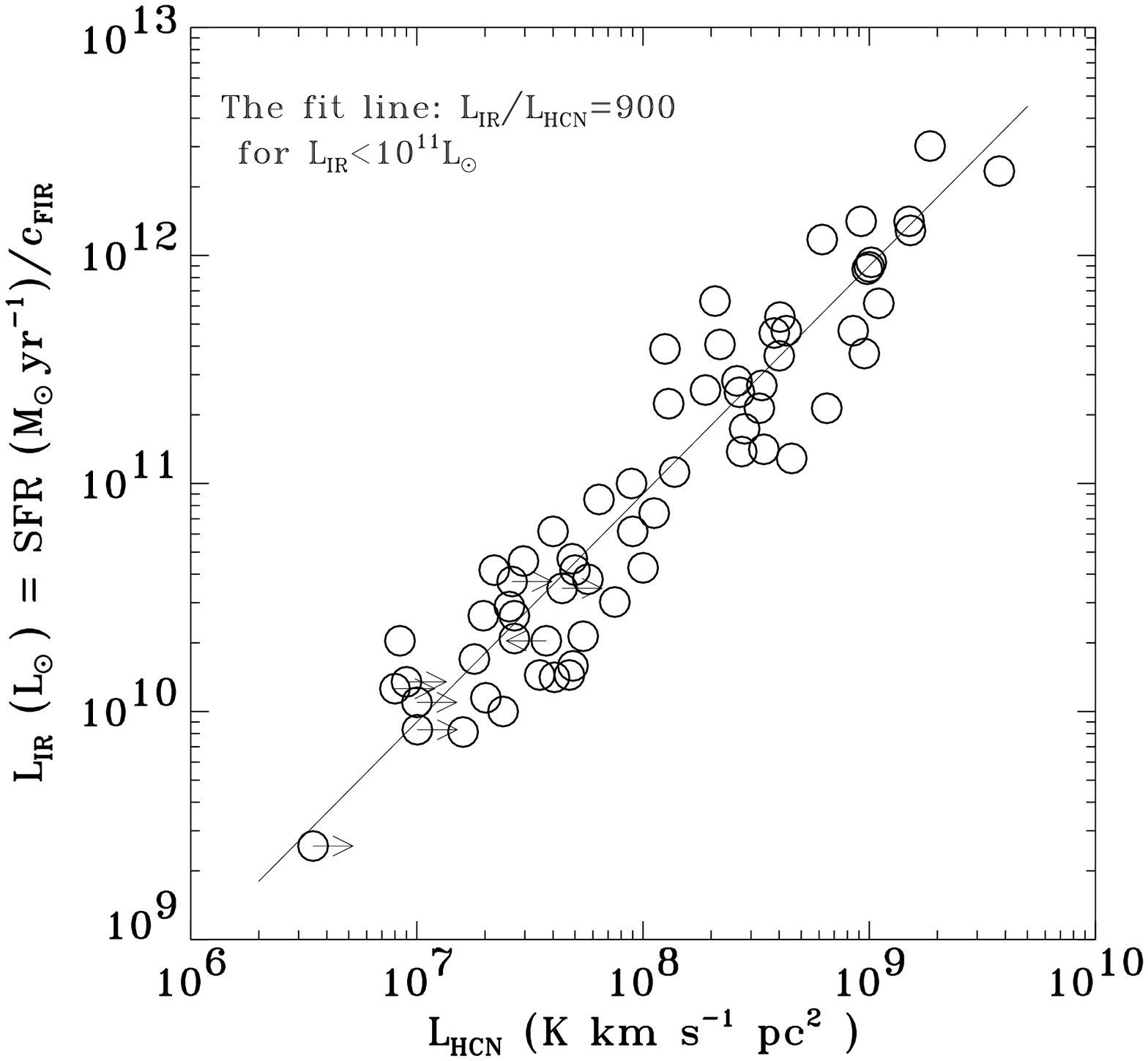}
\vskip -0.4in
\plotone{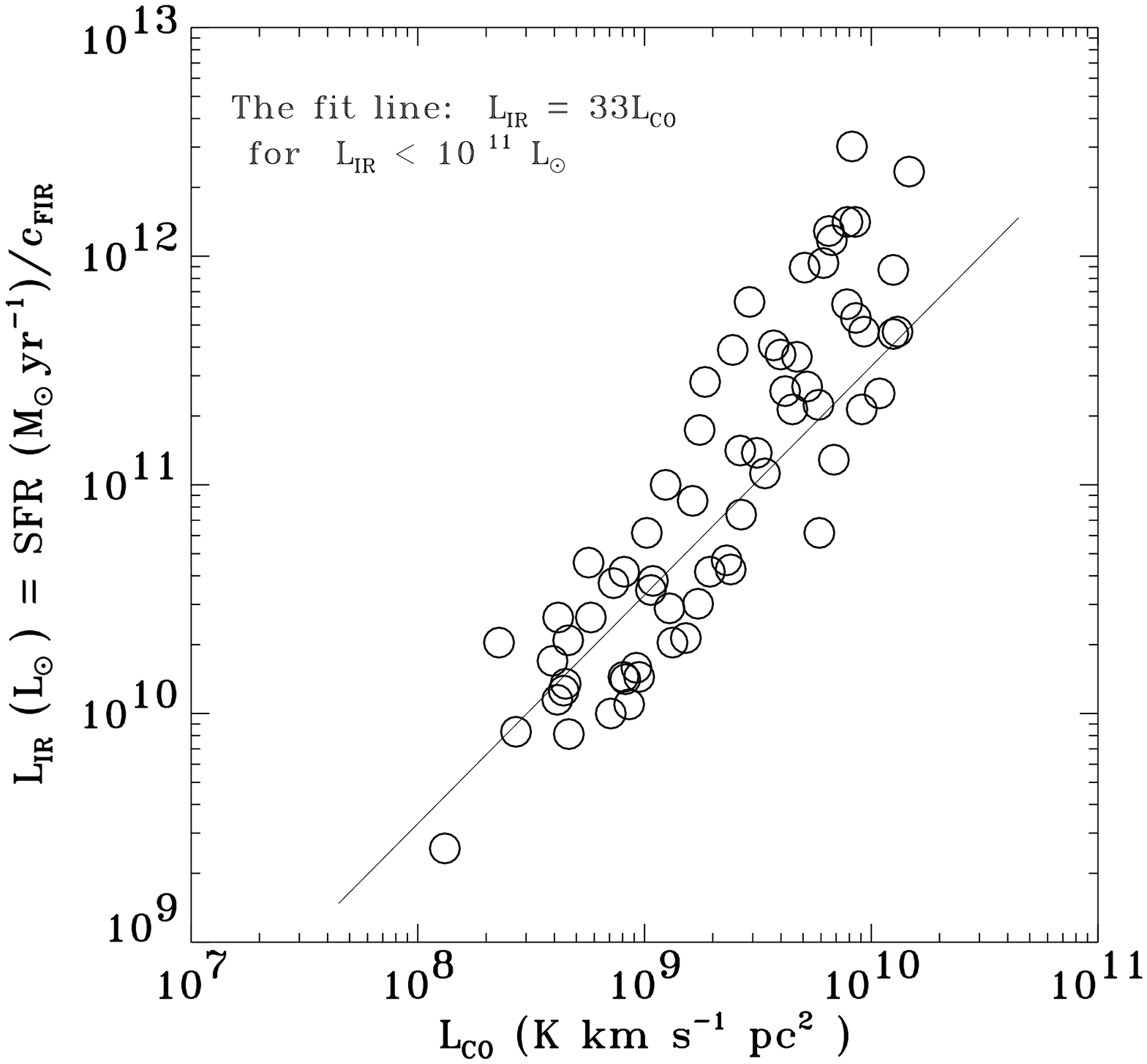}
\vskip -0.2in
\figcaption{(a) The correlation between HCN and IR 
luminosities in 65 galaxies. Some limits in HCN luminosities 
are indicated with arrows. (b) The correlation between $L_{\rm CO}$ 
and $L_{\rm IR}$ for the same HCN sample.
The sample is divided into luminous infrared galaxies (LIGs) 
and ultraluminous infrared galaxies (ULIGs) 
with $L_{\rm IR} \ge 10^{11} \ls$ and less luminous ``normal'' 
spiral galaxies. The solid lines are the fits to the less luminous
galaxies with a slope fixed at unity. A single slope of 1.0 fits HCN data for 
both low and high IR luminosities, but not for the CO data.
\label{fig1}} 
\end{figure*}

\newpage

\begin{figure}
\epsscale{0.7}
\vskip -0.5in
\plotone{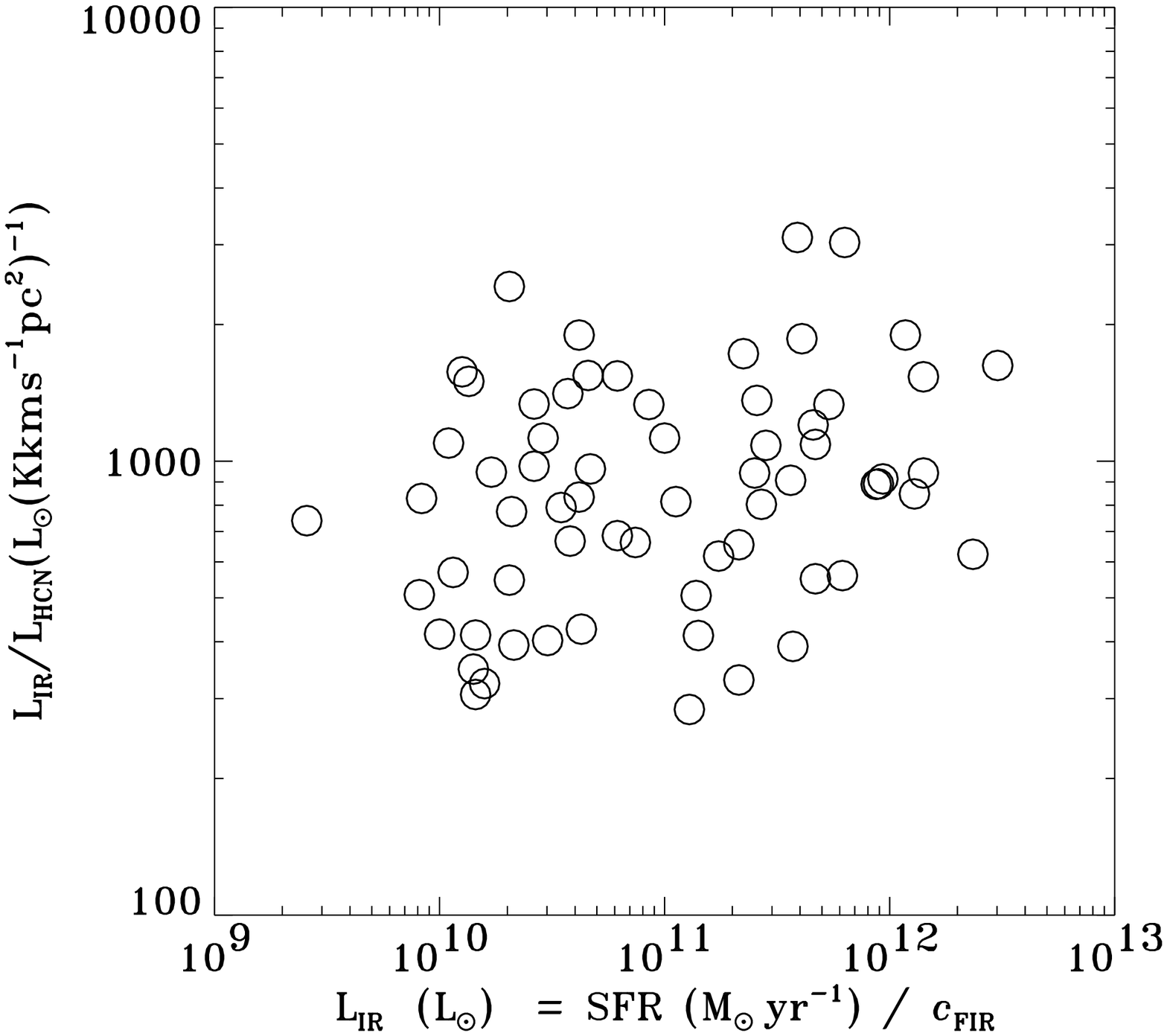}
\vskip -0.4in
\plotone{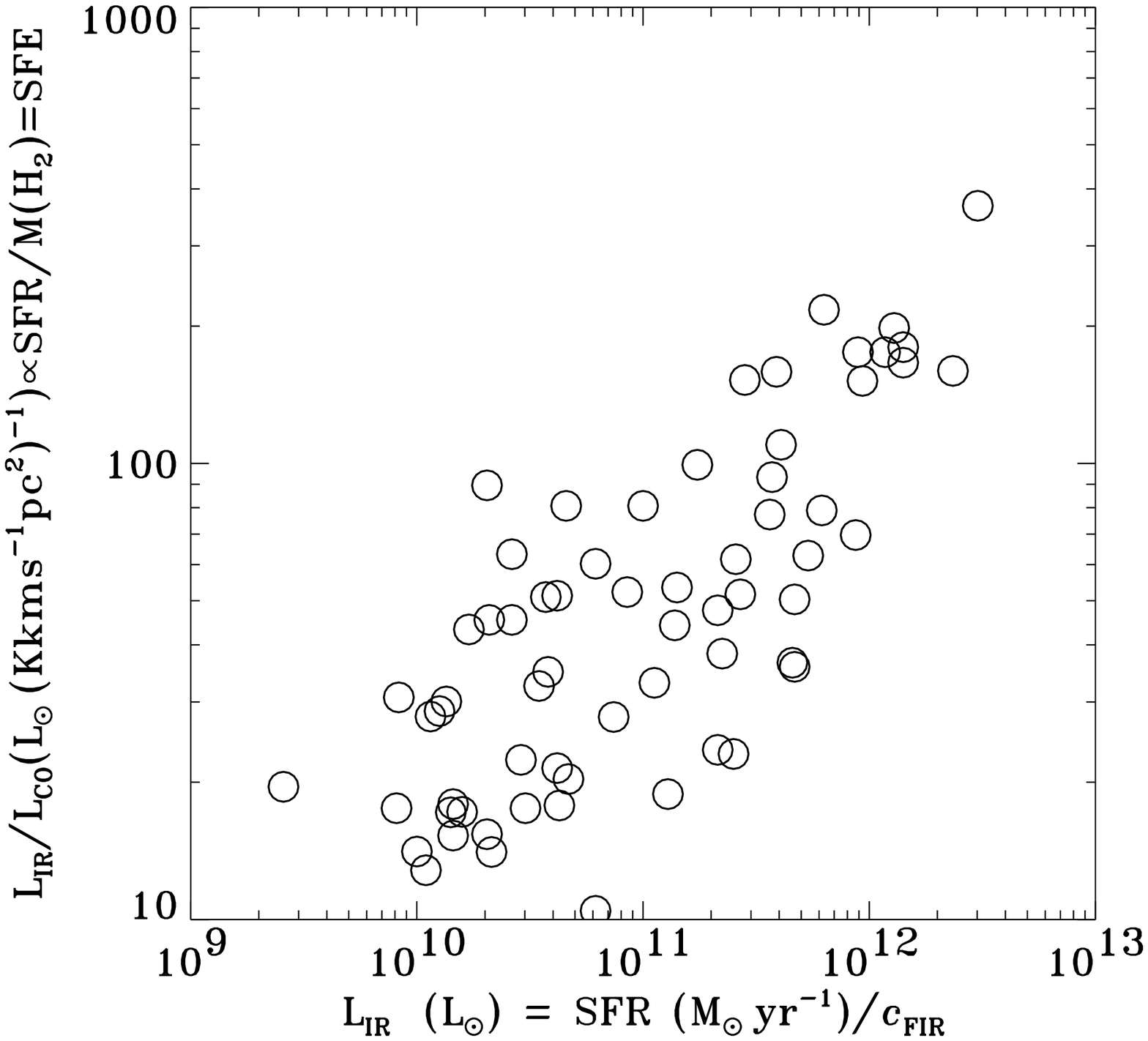}
\vskip -0.2in
\figcaption{(a) The correlation between IR/HCN and IR is
almost of nonexistence; (b) whereas the correlation between 
$L_{\rm IR}/L_{\rm CO}$ and $L_{\rm IR}$ is rather prominent.
The different trends are particularly obvious at high IR luminosity
for LIGs and ULIGs with $L_{\rm IR} \ge 10^{11} \ls$. 
\label{fig2}}
\end{figure}

\newpage
\begin{figure}
\epsscale{.99}
\plotone{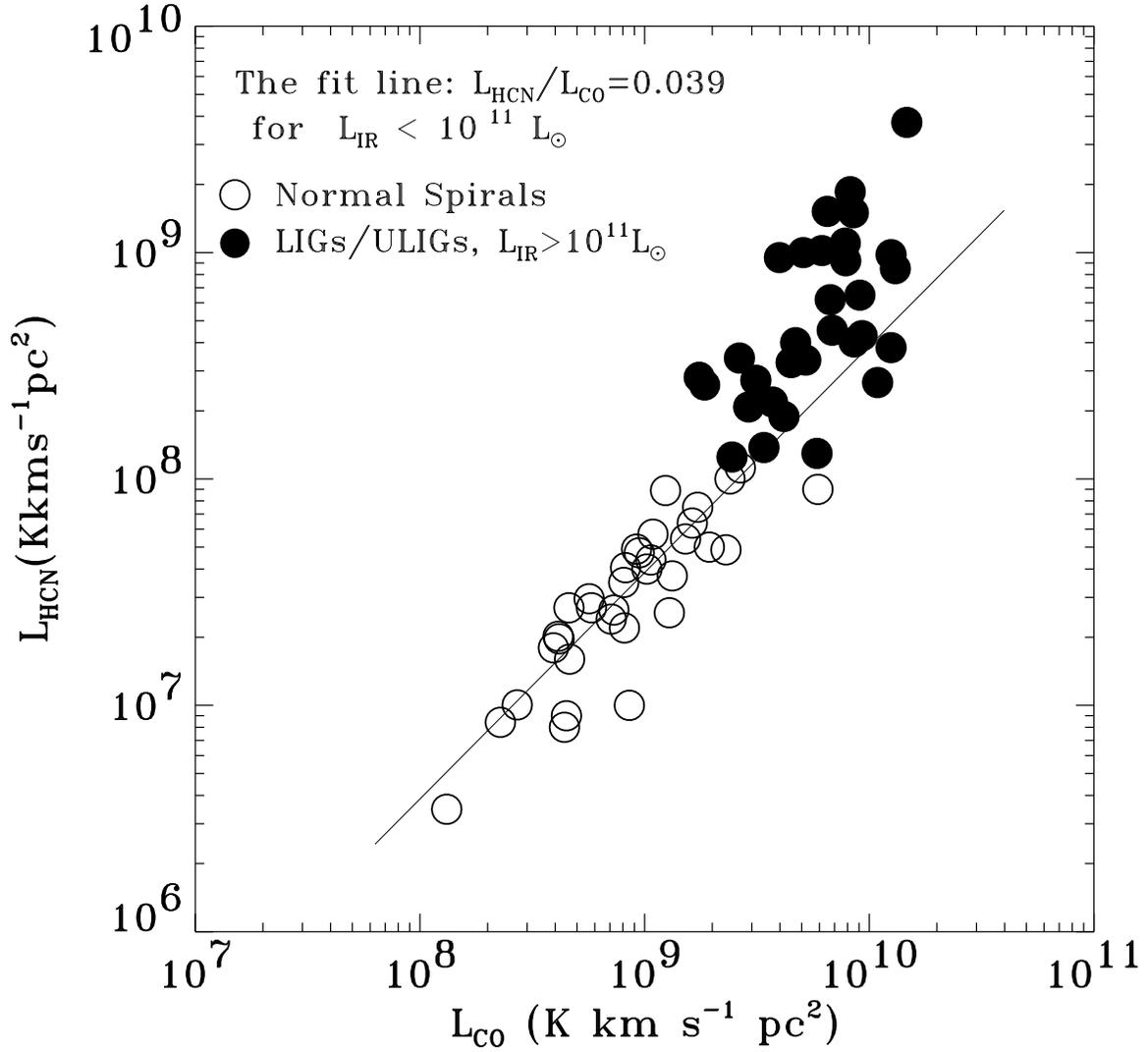}
\figcaption{Strong correlation exists between \lhcn~ and \lco~ indicating
that the more molecular gas-rich galaxies tend to have larger amount of 
dense molecular gas as well. Similar to Figure~1b, the fit line is for
less luminous normal galaxies (open circles) with a fixed slope at 
unity. Apparently, almost all LIGs and ULIGs (filled circles, 
$L_{\rm IR} \ge 10^{11} \ls$) lie above the line.
\label{fig3}}
\end{figure}

\newpage
\begin{figure}
\epsscale{.99}
\plotone{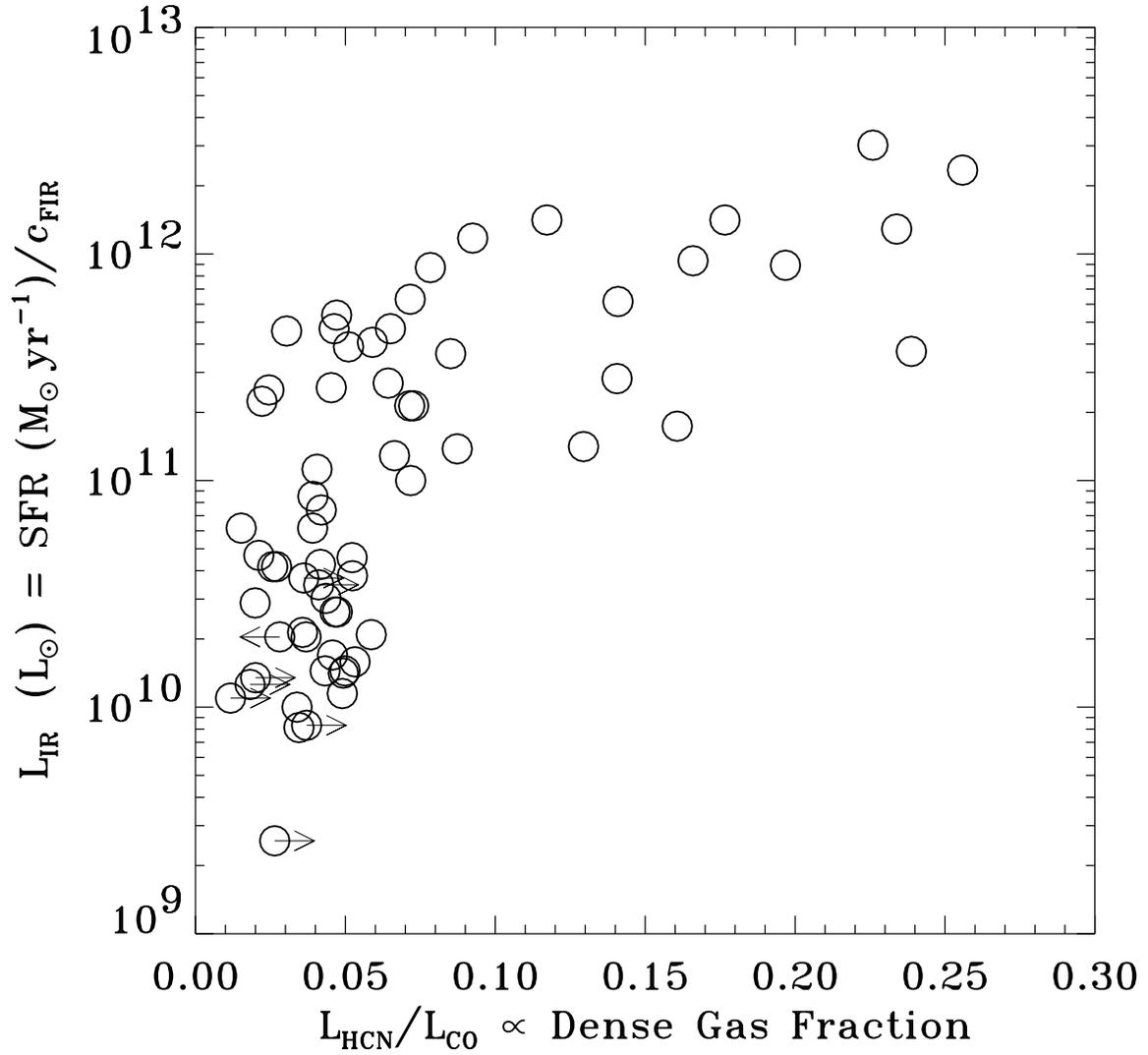}
\figcaption{\lhcn/\lco~ stays at fairly small values for normal 
spiral galaxies, but increases dramatically for LIGs and ULIGs
($L_{\rm IR} \ge 10^{11} \ls$). All galaxies with 
\lhcn/\lco$ \approxgt 0.06$ in the sample are luminous 
and ultraluminous galaxies.
\label{fig4}}
\end{figure}

\begin{figure}
\epsscale{0.7}
\vskip -0.5in
\plotone{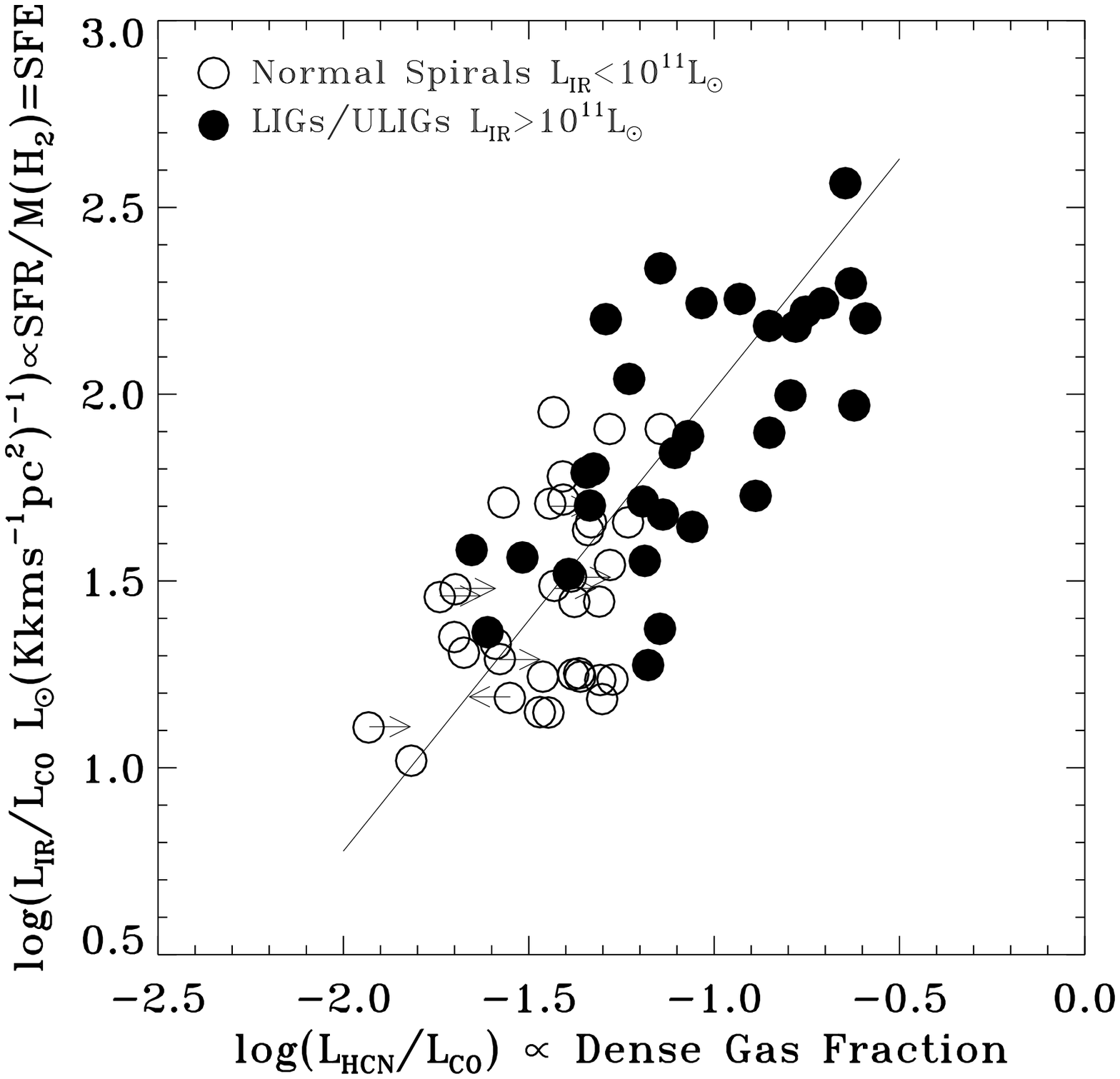}
\vskip -0.4in
\plotone{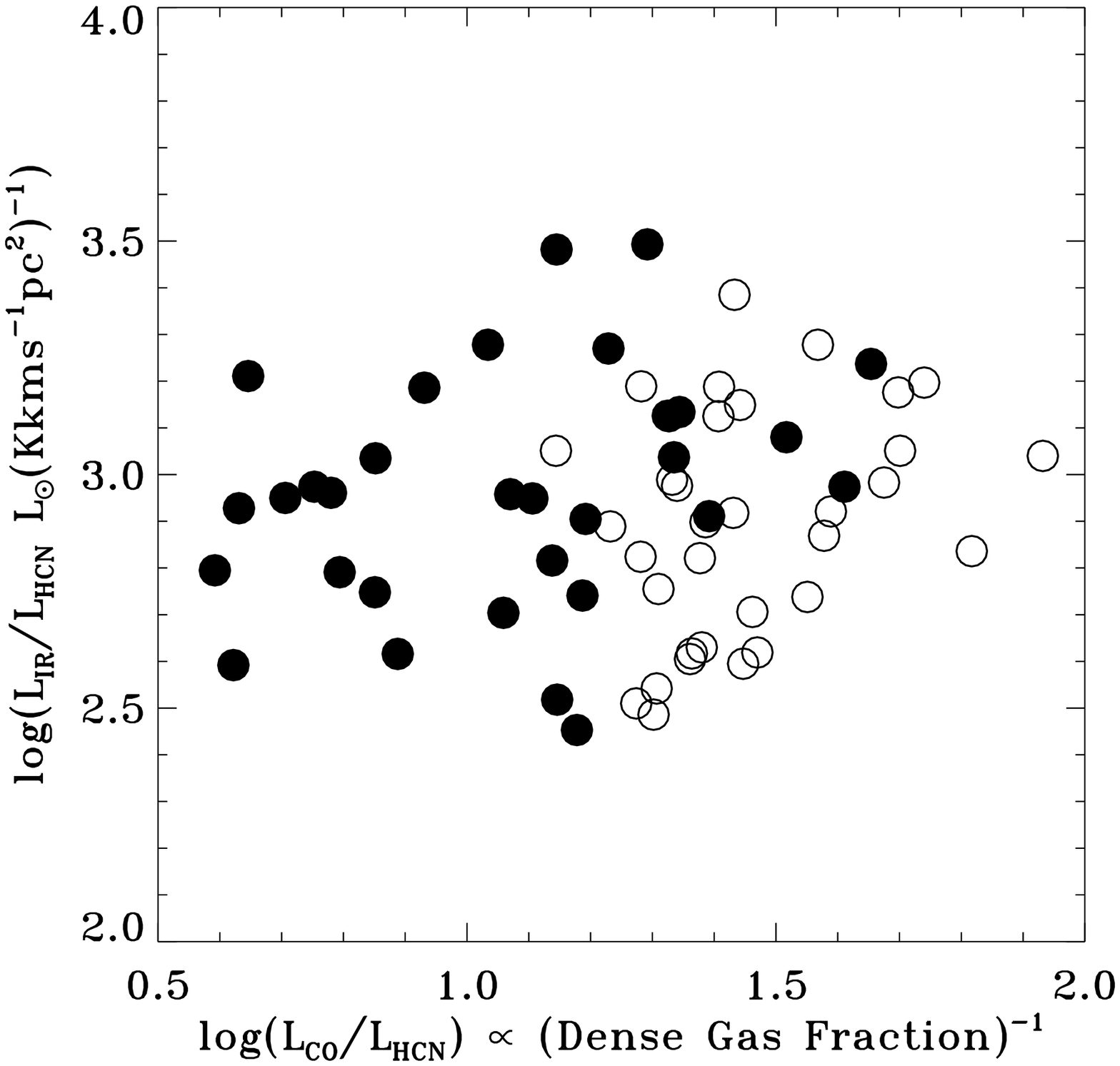}
\vskip -0.3in
\figcaption{ (a) Correlation between $L_{\rm HCN}/L_{\rm CO}$ and $L_{\rm 
IR}/L_{\rm CO}$ revealing the true relationship between the HCN and IR 
since both luminosities are normalized by \lco. This removes all 
dependence on distance and galaxy size and shows that
there is a true physical correlation between the IR and HCN luminosities. 
The best fit has a correlation coefficient of 0.74.
(b) No correlation between $L_{\rm IR}/L_{\rm HCN}$ and 
$L_{\rm CO}/L_{\rm HCN}$ (a correlation coefficient of 0.1) 
suggests that the correlation between $L_{\rm IR}$
and $L_{\rm CO}$ observed in Figure~1b may not be a truly physical relation.
The sample is divided into LIGs and ULIGs with 
$L_{\rm IR} \ge 10^{11} \ls$ (filled circles) and less luminous 
normal spiral galaxies (open circles).
\label{fig5}}
\end{figure}

\newpage
\begin{figure}
\epsscale{.99}
\plotone{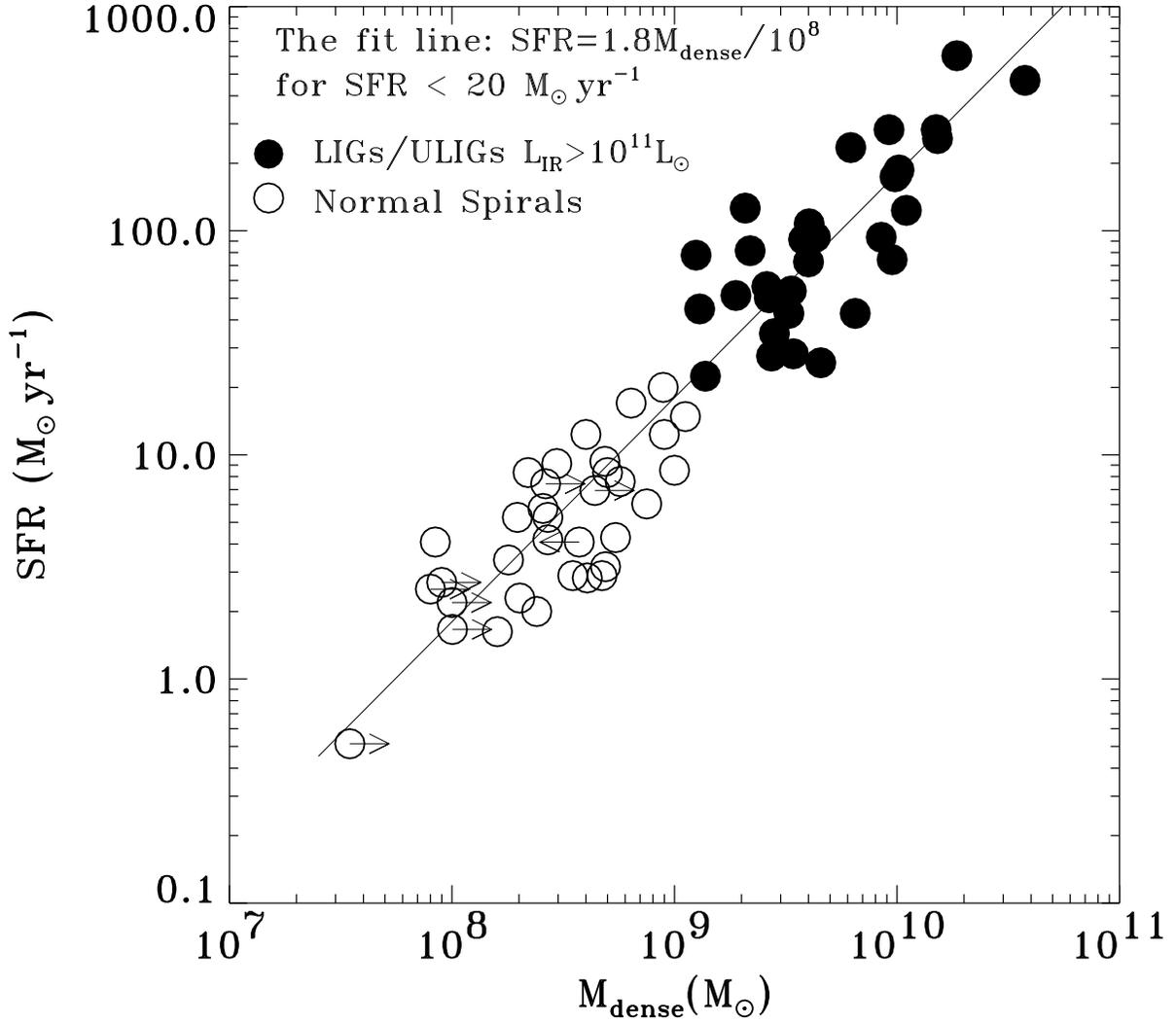}
\figcaption{The global star formation law of dense molecular gas.
The star formation rate is linearly proportional to the dense molecular 
gas mass. The solid circles are for LIGs and ULIGs with
\lir$\approxgt 10^{11}\ls$, whereas the open circles are for the
less luminous normal spiral galaxies. 
\label{fig6}}

\end{figure}

\newpage
\begin{figure}
\epsscale{.99}
\plotone{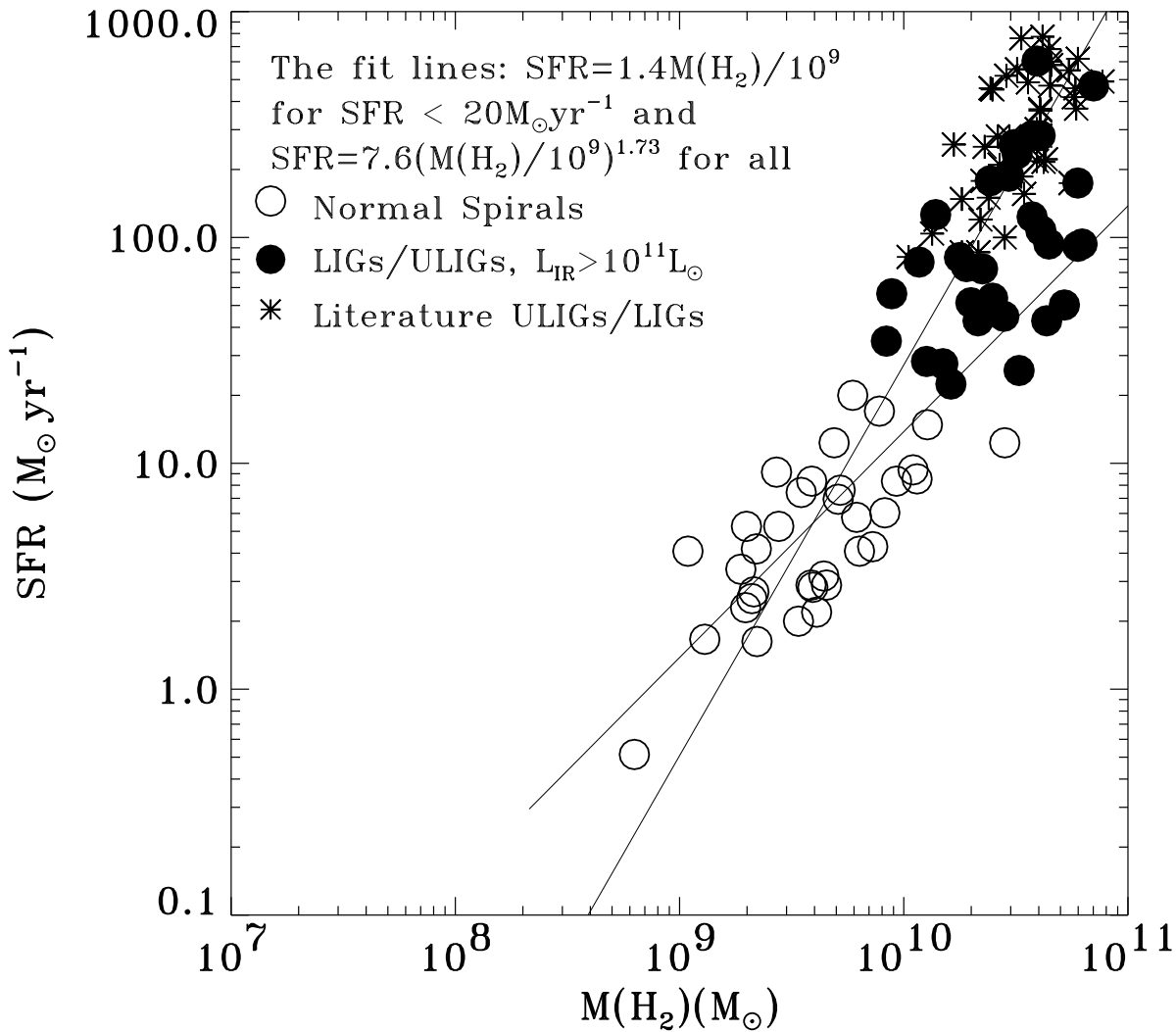}
\figcaption{The star formation rate vs. total molecular gas mass. 
The solid circles are for LIGs and ULIGs with
\lir$\approxgt 10^{11}\ls$, whereas the open circles are for the
less luminous normal spiral galaxies. More CO data of mostly ULIGs
available from literature (stars) are added to our HCN sample 
(see Fig.~1b). The orthogonal least-squares fit now has a slope 
of 1.73 with a correlation coefficient $R=0.89$. The line of fixed slope 
of 1, a valid fit for normal spirals, is also shown for comparison.
\label{fig7}}
\end{figure}

\newpage
\begin{figure}
\epsscale{.99}
\plotone{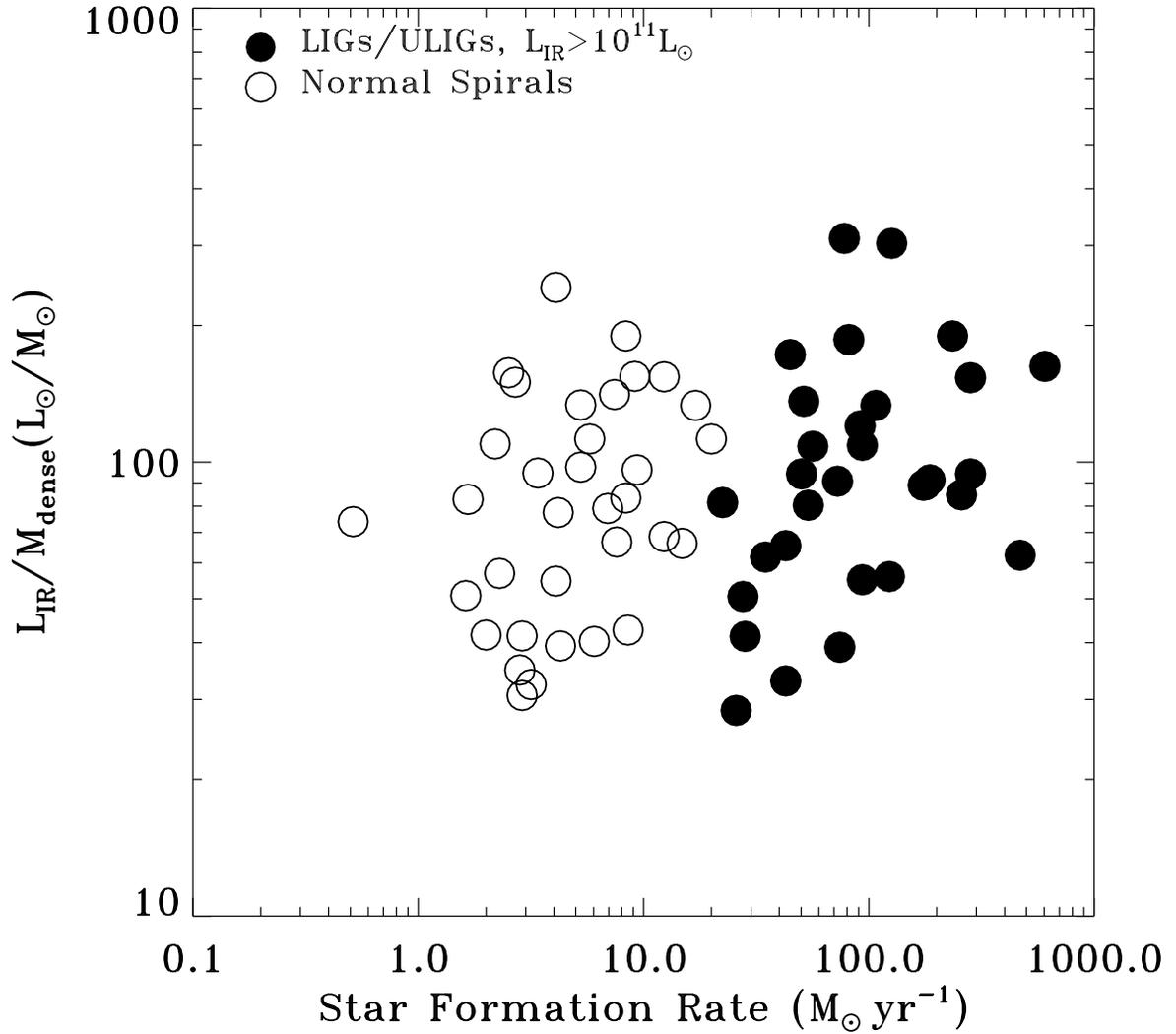}
\figcaption{The star formation rate (far-IR luminosity) per unit 
of {\it dense} molecular gas mass is essentially independent 
of the total IR luminosity. The average for the entire sample 
is \lir/M$_{\rm dense}=90\ls/\ms$. 
\label{fig8}}
\end{figure}

\clearpage

\appendix

\section{Correlations with the Warm Dust Temperature}

The temperature $T_{\rm dust}$ can be estimated as a function of 
$f_{\rm 60\mu m}/f_{\rm 100\mu m}$ and $\beta$, where the 
emissivity is the Planck function of a single warm dust temperature
times the frequency to the power of $\beta$, i.e., $\nu^{\beta}$.
This is for warm dust $T_{\rm dust}$ between $\sim 25$ and 60~K and 
$~\beta \sim $~1---2 (\eg, see the appendix of Lonsdale \etal 1985). 
Although there is increasing evidence
for a substantial amount of cold dust in spiral galaxies (\eg, 
Alton \etal 2000), the warm dust is probably still the most 
important component, especially in luminous and ultraluminous
infrared galaxies (LIGs and ULIGs), and often a single dust
temperature fitting of the far-IR spectral energy distribution (SED) 
including submillimeter measurements is still 
a good approximation (\eg, Lisenfeld, Isaak, \& Hills 2000; 
Dunne et al. 2000).

The tightness of the IR--HCN correlation 
can be further examined by comparing the ratio of 
$L_{\rm IR}/L_{\rm HCN}$ with $f_{\rm 60\mu m}/f_{\rm 100\mu m}$.
Figure~9a shows that $L_{\rm IR}/L_{\rm HCN}$ has only a weak dependence 
upon $f_{\rm 60\mu m}/f_{\rm 100\mu m}$ or $T_{\rm dust}$, whereas 
Figure~9b indicates that $L_{\rm IR}/L_{\rm CO}$ correlates strongly 
with $f_{\rm 60\mu m}/f_{\rm 100\mu m}$ ($R^2=0.72$) or $T_{\rm dust}$. 
The fits in terms of dust temperature have been given in equations (6) 
and (7).
Unlike Figure~5b, where $L_{\rm IR}/L_{\rm HCN}$ appears to be
totally independent of 
$L_{\rm CO}/L_{\rm HCN}$, Figure~9a indicates that 
$L_{\rm IR}/L_{\rm HCN}$ might still be weakly dependent 
upon $T_{\rm dust}$. There is also a 
meaningful correlation between $L_{\rm HCN}/L_{\rm CO}$ and 
$f_{\rm 60\mu m}/f_{\rm 100\mu m}$ ($R^2=0.46$). An orthogonal fit 
can be obtained as $\lhcn/\lco=10^{-0.71}\times 
(f_{\rm 60\mu m}/f_{\rm 100\mu m})^{2.0\pm 0.8}$ with a fairly large
uncertainty in the slope since the correlation is not tight. 

It is not surprising that $L_{\rm IR}/L_{\rm CO}$ is a strong 
function of $T_{\rm dust}$ since the dust radiates thermally 
as \ $T_{\rm dust}^4$ \, to \ $T_{\rm dust}^6$, depending on the
dust emissivity to produce IR emission (\eg, Soifer \etal 1989). 
Thus, Figure~9 is not independent of Figure~2 as $T_{\rm dust}$
is closely related to $L_{\rm IR}$. And \lco~ depends only 
linearly on the temperature, $\sim T_{\rm dust}^1$,
if the dust and gas are coupled and the gas is thermalized with 
the intrinsic brightness temperature \ $T_b \sim T_{\rm dust}^1$ \ 
(\eg, the black-body model of Solomon \etal 1997),
and more weakly in the power 
of $T_{\rm dust}$ dependence if not coupled/thermalized. 
HCN luminosity seems also to depend, at a first glance, 
only on the first power of $T_{\rm dust}$ \ if \ $T_b \sim T_{\rm dust}^1$, 
but it is actually not simply proportional 
to $T_{\rm dust}^1$ \ since the HCN emission
does not fill the source area probed by the telescope beam, whereas 
CO has probably an area filling factor close to unity, especially 
for LIGs/ULIGs. 
Recent interferometric HCN imaging in some nearby galaxies
indeed shows much smaller HCN source size than that of CO 
(\eg, Downes \etal 1992; Helfer \& Blitz 1997a; Kohno \etal 1996,
 1999). It is possible that the HCN
filling factor depends on some power of $T_{\rm dust}$.
In addition, the weak correlation between 
$L_{\rm IR}/L_{\rm HCN}$ and $T_{\rm dust}$ indicates that 
there might be other important parameters, probably the molecular
gas density, responsible for the high ~\lhcn~ in LIGs/ULIGs.
Thus, $L_{\rm HCN}$ depends on a much higher power of $T_{\rm dust}$, 
and the ratio of $L_{\rm HCN}/L_{\rm CO}$ should depend fairly strongly
on some power 
of $T_{\rm dust}$ (eq. [8]) rather than being independent. 

This rather strong $T_{\rm dust}$ dependence  of $L_{\rm HCN}$
can be understood since 
$L_{\rm HCN}/L_{\rm CO}$ also correlates with \lir~ (Fig.~4).
Because the \lhcn/\lco~ ratio is such a good indicator of starburst and
most LIGs/ULIGs have the highest \lhcn/\lco~ ratio, 
warmer LIGs/ULIGs of higher $T_{\rm dust}$ (higher \lir) should have a higher 
fraction of the dense molecular gas (higher $L_{\rm HCN}/L_{\rm CO}$), 
and thus very likely high 
average gas density. Therefore, \lir~ somehow correlates 
with the molecular gas density and $T_{\rm dust}$ is thus also 
related to the gas density, which should be expected as the gas
and dust are somehow coupled.

\clearpage

\begin{deluxetable}{lrrrrrrrrr}
\tablenum{3}
\tablecaption{Model Fit Parameters \label{tbl-3}}
\tablehead{
\colhead{Para.}      &   
\colhead{$L_{\rm 60 \mu m}$}   &   
\colhead{$L_{\rm 100 \mu m}$}   &   
\colhead{$L_{\rm IR}$}     &
\colhead{$T_{\rm dust}$}        &   \colhead{$L_{\rm HCN}$ }    &
\colhead{$L_{\rm CO}$}          &   \colhead{Const.}         &
\colhead{RMS\tablenotemark{a}} & \colhead{$R^2$\tablenotemark{b}} }
\footnotesize
\startdata

\cutinhead{3 parameter correlations}

$L_{\rm IR}$ &      &    &    &      &0.88&0.16& 2.29&0.24&0.87\nl

$L_{\rm IR}$ &      &    &    & 5.4  &    &0.97&$-$6.34&0.17&0.94\nl
$L_{\rm IR}$ &      &    &    & 2.9  &0.86&    &$-$0.48&0.20&0.92\nl

$L_{\rm HCN}$ &      &    &    & 3.3  &    &1.00&$-$6.22&0.21&0.89\nl
$L_{\rm HCN}$ &      &    &0.56&      &    &0.49&$-$2.62&0.19&0.90\nl

$L_{\rm HCN}$ &      &    &0.96&($-$1.7)&    &    & 0.17&0.21&0.89\nl  
$L_{\rm CO} $ &      &    &0.89&$-$4.4  &    &    & 6.27&0.17&0.88\nl  

$L_{\rm HCN}$ &      &0.95&    &($-$0.2)&    &    &$-$1.12&0.21&0.89\nl  
$L_{\rm CO} $ &      &0.88&    &$-$3.0  &    &    & 5.07&0.17&0.88\nl  

$L_{\rm HCN}$ &($-$.06)&1.01&    &      &    &    &$-$1.41&0.21&0.89\nl  
$L_{\rm CO} $ &$-$1.13 &2.02&    &      &    &    & 0.47&0.17&0.88\nl  
  
\cutinhead{simple 2 parameter correlations}

$L_{\rm IR}$ &      &    &    &      &    &1.25& 0.72&0.34&0.77\nl
$L_{\rm IR}$ &      &    &    &      &1.02&    & 2.58&0.24&0.88\nl

$L_{\rm HCN}$ &      &    &    &      &    &1.26&$-$3.73&0.28&0.85\nl  
$L_{\rm HCN}$ &      &    &0.91&      &    &    &$-$1.93&0.23&0.87\nl  
$L_{\rm HCN}$ &      &0.96&    &      &    &    &$-$1.52&0.22&0.89\nl  

$L_{\rm CO}$  &      &0.72&    &      &    &    &$-$2.12 &0.22&0.80\nl  
$L_{\rm CO}$  &      &    &0.62&      &    &    &2.49 &0.30&0.75\nl  
$L_{\rm CO}$  &      &    &    &      &0.67&    &3.92&0.27&0.85\nl  
$M_{\rm dust}$ &     &    &    &      &0.84&    &0.41&0.25&0.84\nl
$M_{\rm dust}$ &     &    &    &      &    &1.13& $-$3.35  &0.21&0.87\nl

\tablenotetext{a}{rms $=$ Root Mean Square deviation (logarithmic) between 
the data (observed) and the predicted from the (correlation) model fits.}
\tablenotetext{b}{The squared correlation coefficient $R^2$.}
\tablecomments{All quantities are logarithmic in the correlation fits. The 
line luminosities are in units of ${\rm K \kms pc}^2$ and the total IR,
60, and 100 $\mu$m luminosities are in units of \ls, while $T_{\rm dust}$
is in K. The warm dust mass $M_{\rm dust}$ was estimated using only 
the 100~$\mu$m flux and warm dust temperature.
The model fit for $L_{\rm IR}$ from $L_{\rm HCN}$ and 
$L_{\rm CO}$ (the IR(\lhcn, \lco) model), for example, is 
$logL_{\rm IR}(predicted)=0.88logL_{\rm HCN}+0.16logL_{\rm CO}+2.3$
with the correlation coefficient $R=0.93$ ($R^2=0.87$) and the rms error is 
$(\Sigma[logL_{\rm IR}(predicted)-logL_{\rm IR}(obs.)]^2/N)^{1\over 2}=0.24$.
Numbers with parentheses indicate only marginal significance.}

\enddata
\end{deluxetable}

\section{Other Correlations and Model Fits}

\subsection{Simple Two-parameter Fits}

Although ~\lhcn~ and \lco~ are strongly correlated (Fig.~3), there are 
significant differences, however, when they are
compared with either $L_{\rm 100 \mu m}$ or $L_{\rm IR}$ (Table~B1).
The better correlation between $L_{\rm HCN}$ and $L_{\rm IR}$
(vs $L_{\rm CO}$ and $L_{\rm IR}$)
can be recognized in both $R^2$ (\S 3.1) and the
R.M.S. deviations of the correlation fits (Table~3, 0.24 vs 0.34). 
The luminosity $L_{\rm 100 \mu m}$ also
correlates better with $L_{\rm HCN}$ than with $L_{\rm CO}$
as the difference in $R^2$ is also significant (0.89 vs 0.80). 
Nevertheless, \lco~ appears to be much better correlated with 
$L_{\rm 100 \mu m}$ than with $L_{\rm IR}$ (there is 
a significant difference in the logarithmic R.M.S. deviations: 
0.22 vs 0.34), though the difference in $R^2$
(0.80 vs 0.77) is small. In comparison, there is little difference 
in the correlations between $L_{\rm 100 \mu m}$ and \lhcn~ 
vs between $L_{\rm IR}$ and ~\lhcn~ ($R^2$=0.89 vs 0.87 and logarithmic 
R.M.S. 0.22 vs 0.23). But the HCN correlations are much tighter 
than the CO correlations,
even though the ~$L_{\rm 100 \mu m}$--\lco~ correlation has same R.M.S.
deviation.

The luminosity $L_{\rm 100 \mu m}$ traces relatively the cooler 
component of the 
warm dust emission at $T_{\rm dust} \sim 25$--50 K. In this regime 
the 100~$\mu$m emission is a reasonable tracer of the warm dust mass
$M_{\rm dust}$. In comparison, the total IR luminosity \lir~
contains the mixture of the various dust components including
the hot dust radiated at the mid-IR emission.
We here estimate the warm dust mass $M_{\rm dust}$ from the 100~$\mu$m 
flux density and the warm dust temperature $T_{\rm dust}$ and 
correlate it with \lhcn~ and \lco~ in Table~3 as well. It appears
that \lco~ is only slightly better correlated with $M_{\rm dust}$
than \lhcn~ with $M_{\rm dust}$. In short, the differences in the 
above-mentioned 
various correlations suggest that HCN is a much better tracer of 
both the total IR and 100~$\mu$m emission than CO, whereas 
CO only traces better the 100~$\mu$m than the total IR emission. 
The dense molecular gas rather than the total molecular gas
is more intimately related to both the total IR and 
100 $\mu$m emission.

\subsection{Multi-parameter Fits}

The predicted \lir~
from the CO and $T_{\rm dust}$ [the IR(\lco, $T_{\rm dust}$) model,
$logL_{\rm IR}(L_{\rm CO},~T_{\rm dust},~model\_fit)
=-6.34+0.97logL_{\rm CO}+5.4logT_{\rm dust}$] turns out to be 
the tightest correlation ($R^2=0.94$) among all. The prediction of \lir~ 
from CO and $T_{\rm dust}$ is more accurate than from HCN alone 
and also slightly better than from HCN plus $T_{\rm dust}$. 
This surely suggests the importance of $T_{\rm dust}$ when the 
total molecular gas content, rather than the dense molecular gas, 
is concerned in predicting \lir. This also 
implies that $T_{\rm dust}$ is 
an important parameter in regulating star formation of the total 
molecular gas, but not necessarily the dense (active star-forming) 
molecular gas.

We also use $T_{\rm dust}$ as well as $L_{\rm 100\mu m}$ and 
$L_{\rm 60\mu m}$, though they are not independent, in the 
three-parameter model fits as listed in Table~3, and we find
the following:

1. The value of \lco~ decreases with $T_{\rm dust}$ for given 
$L_{\rm 100\mu m}$ or $L_{\rm IR}$.
The value of \lhcn~ is, however, linearly proportional to $L_{\rm 100\mu m}$ 
or $L_{\rm IR}$ with no correction or only marginal correction of 
$T_{\rm dust}$. Thus, \lhcn~ decreases little with $T_{\rm dust}$ for a given 
$L_{\rm 100\mu m}$ or $L_{\rm IR}$. To put it in a different perspective, 
\lir~ is only slightly better predicted when $T_{\rm dust}$ is 
considered together with \lhcn, whereas \lir~ is much better predicted 
when both \lco~ and $T_{\rm dust}$ are in the fit.

2. The HCN(\lco, $T_{\rm dust}$) model implies that HCN correlates 
with $T_{\rm dust}$ to the 3.3 power as well as linearly (the 
first power) with \lco~ (cf. Equation 8). Thus, HCN
responds sensitively to $T_{\rm dust}$ with a power of $\approxgt 4.3$.
The IR(\lhcn, $T_{\rm dust}$)
model gives a 2.9 power correlation with $T_{\rm dust}$
and a 0.86 power with \lhcn, again roughly consistent with the 
high ($\sim 6$) power $T_{\rm dust}$ dependence 
(cf. eq. [6]) when the $T_{\rm dust}$ dependence of 
\lhcn~ (eq. [8]) has been considered. In short, the difference in the 
dependences of the HCN and IR upon the warm dust temperature is
not dramatic. Thus, the ratio of \lir/\lhcn~ has only a weak 
dependence upon $T_{\rm dust}$ (eq. [7] and Fig.~9a).

3. The HCN($L_{60 \mu m}$, $L_{100 \mu m}$) model fit
and CO($L_{60 \mu m}$, $L_{100 \mu m}$) 3 parameter correlation
model reveal that HCN correlates primarily with $L_{100 \mu m}$, and 
essentially not at all with $L_{60 \mu m}$, whereas CO correlates with
both $L_{60 \mu m}$ and $L_{100 \mu m}$. This again implies
the much tighter dependence of CO upon $T_{\rm dust}$, and
is basically reiterating 
the first point described above since there are only 
two independent parameters among $T_{\rm dust}$, $L_{60 \mu m}$, 
$L_{100 \mu m}$ and \lir. 

\subsection{Fits Between the Ratios}

Figures.\,4 and 5  already showed some correlations between the parameter 
ratios. Here we just list the weak correlation 
between \lir/\lhcn~ and $f_{\rm 60\mu m}/f_{\rm 100\mu m}$ 
($R^2$=0.17, Fig.~9a), as directly obtained from the orthogonal fit,
$\lir/\lhcn=10^{3.1}\times (f_{60 \mu m}/f_{100 \mu m})^{0.63}$
with a logarithmic R.M.S. of 0.21. 
The $T_{\rm dust}$ power and the fit have large scatters owing to 
the poor correlation. The scatter is about the 
same or only slightly better than the R.M.S. of the observed mean 
luminosity ratio (in logarithm) of $log<$\lir/\lhcn$>$ for 
the entire sample, which is $\sim 0.23$.
The correlation fit for \lir/\lco vs $f_{\rm 60\mu m}/f_{\rm 100\mu m}$
or $T_{\rm dust}$ (Fig.~9b, eq. [6]) is significantly
tighter and has a logarithmic R.M.S. of 0.17. 
This small R.M.S. is obviously much
better than the R.M.S. of the observed mean $log<$\lir/\lco$>$ 
of the entire sample, which is $\sim 0.37$. 

\clearpage

\begin{figure}
\epsscale{0.7}
\vskip -0.55in
 \plotone{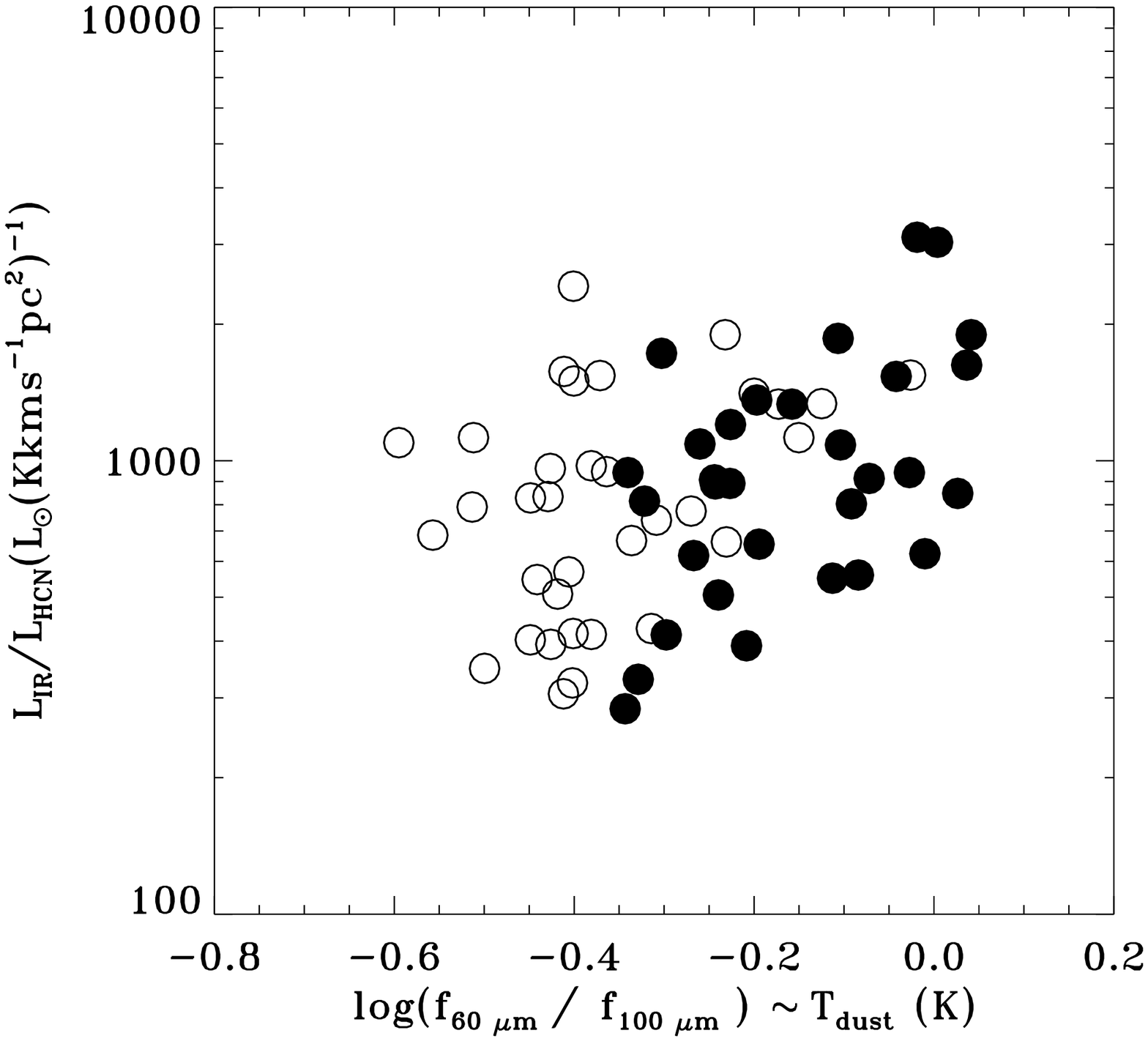}
\vskip -0.35in
\plotone{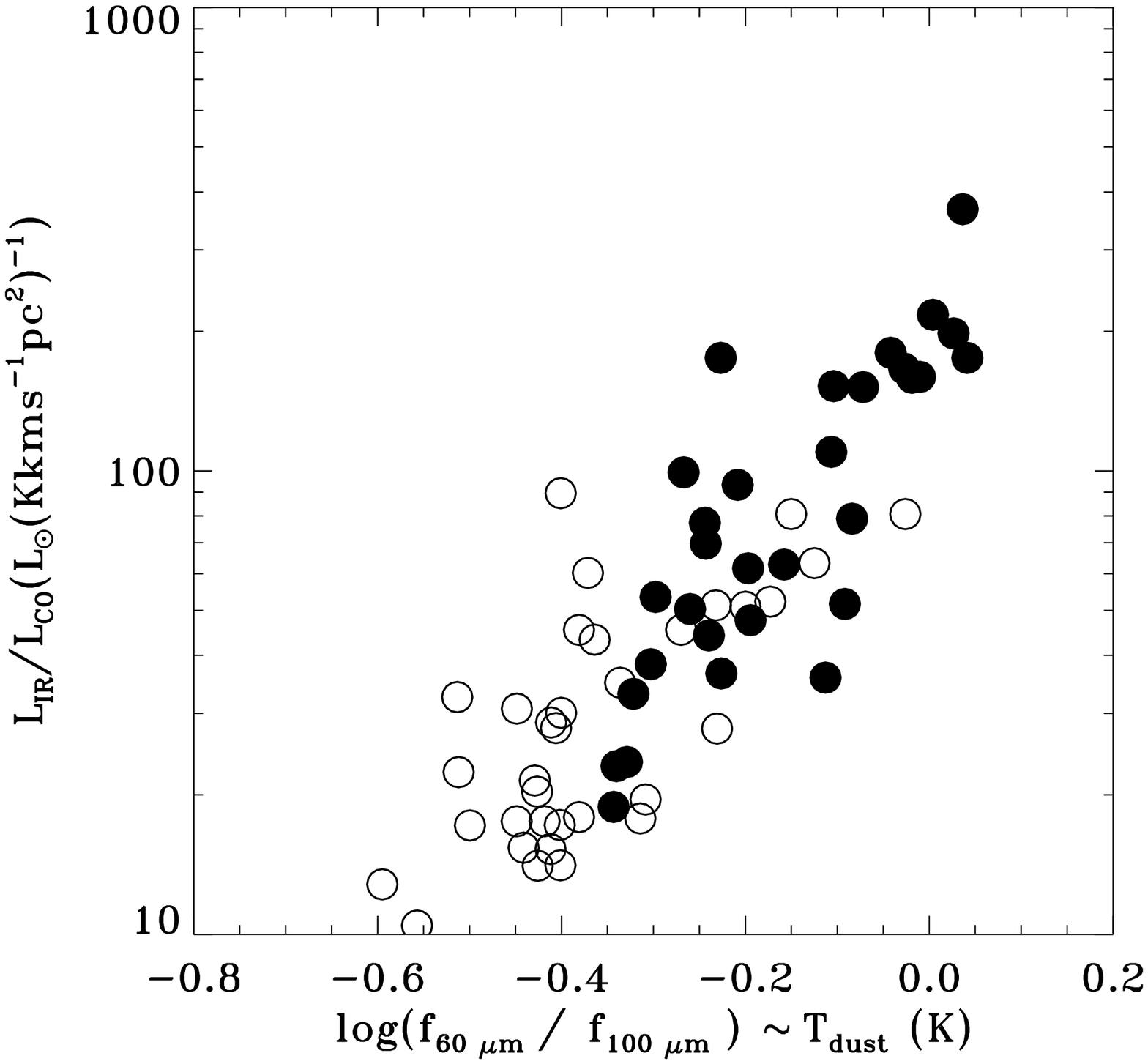}
\vskip -0.2in
\figcaption{(a) $L_{\rm IR}/L_{\rm HCN}$ as a function of the far-infrared
(FIR) color (the 60$\mu$m-to-100$\mu$m flux ratio) which indicates 
the warm dust temperature $T_{\rm dust}$. (b) $L_{\rm IR}/L_{\rm CO}$ as a 
function of the FIR color (or $T_{\rm dust}$).
There is only weak correlation between 
$L_{\rm IR}/L_{\rm HCN}$ and $T_{\rm dust}$, whereas strong 
correlation exists between $L_{\rm IR}/L_{\rm CO}$ and $T_{\rm dust}$.
For optically thin dust with emissivity $\propto \nu^{1.5}$, $T_{\rm dust}$
is approximately in a range of 25--50 K for galaxies in our sample.
The sample is divided into luminous and ultraluminous infrared galaxies
(LIGs/ULIGs) with $L_{\rm IR} \ge 10^{11} \ls$ 
(filled circles) and less luminous galaxies (open circles).
\label{fig-9}}
\end{figure}


\clearpage


\begin{references}

\reference{Aal95} Aalto, S., Booth, R.S., Black, J.H. \& Johansson,
L.E.B. 1995, A\&A, 300, 369
\reference{Aal97} Aalto, S., Radford, S.J.E., Scoville, N.Z., \& 
Sargent, A.I. 1997, ApJ, 475, L107
\reference{} Alton, P.B., Trewhella, M., Davies, J.I., Evans, R., 
Bianchi, S., Gear, W., Thronson, H., Valentijn, E., Witt, A. 1998a, 
A\&A, 335, 807
\reference{} Alton, P.B., Bianchi, S., Rand, R.J., Xilouris, E. 
 M., Davies, J.I., \& Trewhella, M. 1998b, ApJ, 507, L125
\reference{} Alton, P.B., Xilouris, E.M., Bianchi, S., Davies, J., \&
Kylafis, N. 2000, A\&A, 356, 795
\reference{Blo86} Bloeman, J.B.G.M., et al. 1986  A\&A, 154, 25
\reference{} Bryant, P.M., \& Scoville, N.Z. 1999, AJ, 117, 2632
\reference{} Carilli, C.L., Cox, P., Bertoldi, F. \etal 2002, ApJ, 575, 145
\reference{} Casoli, F., Dupraz, C., \& Combes, F. 1992, \aap, 264, 49
\reference{} Casoli, F., Willaime, M.-C., Viallefond, F., \& Gerin, M. 
1999, A\&A, 346, 663
\reference{} Chapman, S.C., Blain, A.W., Ivison, R.J., \& Smail, I. 
2003, Nature, 422, 695
\reference{Con91} Condon, J.J., Huang, Z.-P., Yin, Q.F., \& Thuan, T.X.
1991, ApJ, 378, 65
\reference{cox02} Cox, P. Omont, A., Djorgovski, S.G. \etal 2002, A\&A, 387, 406 (astro-ph/0203355)
\reference{cur00} Curran, S.J., Aalto, S., \& Booth, R.S. 2000, \aaps, 141, 193
\reference{cur01} Curran, S.J., Polatidis, A. G., Aalto, S., \& Booth, R.S. 
2001, \aap, 368, 824
\reference{} Davies, J.I., Alton, P., Trewhella, M., Evans, R., 
\& Bianchi, S. 1999, MNRAS, 304, 495
\reference{Dev90} Devereux, N.A., \& Young, J.S. 1990, ApJ, 350, L25
\reference{Dev92} Devereux, N.A., \& Young, J.S. 1992, AJ,  103, 1536
\reference{Dev93} Devereux, N.A., \& Young, J.S. 1993, AJ,  106, 948
\reference{Dow92} Downes, D., Radford, S.J.E., Guilloteau, S., \etal 1992, 
A\&A, 262, 424
\reference{Dow98} Downes, D., \& Solomon, P.M. 1998, ApJ, 507, 615
\reference{Dow99} Downes, D., Neri, R., Wiklind, T., Wilner, D.J., \& 
Shaver, P.A. 1999, ApJ, 513, L1 
\reference{} Dunne, L., Eales, S., Edmunds, M., Ivison, R., 
Alexander, P., \& Clements, D.L. 2000, MNRAS, 315, 115
\reference{} Evans, N.J., II 1999, ARA\&A, 37, 311 
\reference{} Franceschini, A., Braito, V., Persic, M. \etal 2003, MNRAS, 
343, 1181
\reference{} Frayer, D.T. \etal 1998, ApJ, 506, L7
\reference{Gao96} Gao, Y. 1996, Ph.D. thesis, SUNY at Stony Brook 
\reference{Gao97} Gao, Y. 1997, PASP, 109, 1189 
\reference{Gao97} Gao, Y., Solomon, P.M., Downes, D., \& Radford, S.J.E. 1997,
ApJ, 481, L35 
\reference{GS99} Gao, Y., \& Solomon, P.M. 1999, ApJ, 512, L99
\reference{Gao99} Gao, Y., Gruendl, R.A., Hwang, C.-Y., \& Lo, K.Y. 
1999, in IAU 186, 227
\reference{Gao03} Gao, Y., \& Solomon, P.M. 2003, ApJ, submitted (Paper I)
\reference{ } Gallagher, J.S., \& Hunter, D.A. 1987, in {\it Star Formation
in Galaxies,} ed. C. Lonsdale, (Washington: Gov't Ptg Office), p. 167
\reference{} Gehrz, R.D., Sramek, R.A., \& Weedman, D.W. 1983, 267, 551
\reference{} Genzel, R., et al 1998, ApJ, 498, 579
\reference{} Greve, T.R., Ivison, R.J., \& Papadopoulos, P.P. 
2003, ApJ, 599, 839 (astro-ph/0309213)
\reference{} Guilloteau, S. \etal 1999, \aap, 349, 363
\reference{} Haas, M., Lemke, D., Stickel, M. \etal 1998, A\&A, 338, 33
\reference{} Heckman, T.M. 2000, Philos.Trans.R.Soc.London, A, 358, 2077
\reference{Hel93} Helfer, T.T., \& Blitz, L.  1993, ApJ, 419, 86 
\reference{Hel95} Helfer, T.T., \& Blitz, L.  1995, ApJ,  450, 90 
\reference{Hel97} Helfer, T.T., \& Blitz, L.  1997a, ApJ,  478, 162 
\reference{HB97b} Helfer, T.T., \& Blitz, L.  1997b, ApJ,  478, 233 
\reference{Hen90} Henkel, C., Whiteoak, J.B., Nyman, L.-\AA., \& Harju,
J. 1990, A\&A, 230, L5
\reference{Hen94} Henkel, C., Whiteoak, J.B., \& Mauersberger, R. 1994, 
A\&A, 284, 17
\reference{} Hwang, C.-Y., Lo, K.Y., Gao, Y., Gruendl, R.A., \& Lu, N.Y. 1999,
ApJ, 511, L17
\reference{} Isaak, K.G., \etal 2002, MNRAS, 329, 149
\reference{Isr92} Israel, F.P. 1992, A\&A, 265, 487
\reference{Jac93} Jackson, J.M., Paglione, T.A.D., Ishizuki, S., \&
Nguyen-Q-Rieu 1993, ApJ, 418, L13
\reference{Jac96} Jackson, J.M., Heyer, M.H., Paglione, T.A.D., \& 
Bolatto, A.D. 1996, ApJ, 456, L91
\reference{} Joseph, R. 1999, Ap\&SS, 266, 321
\reference{} Kennicutt, R.C. 1998a, \apj, 498, 541
\reference{} Kennicutt, R.C. 1998b, \araa, 36, 189
\reference{} Kohno, K., \etal 2003, PASJ, 55, L1
\reference{} Kohno, K., Kawabe, R., \& Vila-Vilaro, B. 1999, ApJ, 511, 157
\reference{} Kohno, K., Kawabe, R., Tosaki, T., \& Okumura, S.K. 1996, 
ApJ, 461, L29
\reference{} Lee, Y., Snell, R.L., \& Dickman, R.L. 1990, ApJ, 355, 536
\reference{} Lisenfeld, U., Isaak, K.G., \& Hills, R. 2000, MNRAS, 312, 433 
\reference{} Lo, K.Y., Gao, Y., \& Gruendl, R.A. 1997, 475, L103
\reference{} Lonsdale, C.J., Helou, G., Good, J.C., \& Rice, W. 1985,
{\it Cataloged galaxies and quasars observed in the IRAS survey} (Pasadena: JPL)
\reference{} Luhman, M.L., Satyapal, S., Fischer, J., Wolfire, M.G., 
Sturm, E., Dudley, C.C., Lutz, D., \& Genzel, R. 2003, \apj, 594, 758 
\reference{MH93} Mauersberger, R., \& Henkel, C. 1993, Rev. Modern Astron., 
6, 69
\reference{mt01} Meier, D.S., \& Turner, J.L. 2001, ApJ, 551, 687
\reference{mir90} Mirabel, I.F., \etal 1990, A\&A, 236, 327
\reference{ms88} Mooney, T.J., \& Solomon, P.M. 1988, ApJ, 334, L51
\reference{} Nagar, N.M., ilson, A.S., Falcke, H., Veilleux, S., \&
 Maiolino, R. 2003, A\&A, 409, 115 (astro-ph/0309298)
\reference{} Neri, R., \etal 2003, ApJ, in press (astro-ph/0307310)
\reference{Ngu89} Nguyen-Q-Rieu, Nakai, N. \& Jackson, J.M. 1989, A\&A, 
220, 57
\reference{Ngu92} Nguyen-Q-Rieu, Jackson, J.M., Henkel, C., Truon-Bach 
\& Mauersberger, R. 1992, ApJ, 399, 521
\reference{} Omont, A., \etal 2001, \aap, 374, 371
\reference{Pag95} Paglione, T.A.D., Tosaki, T., \& Jackson, J.M. 1995, 
ApJ, 454, L117 
\reference{Pag97} Paglione, T.A.D., Jackson, J.M., \& Ishizuki, S. 1997, 
ApJ, 484, 656 
\reference{Pap01} Papadopoulos, P.P., Ivison, R., Carilli., C., \& Lewis, G. 
2001, Nature, 409, 58
\reference{pir99} Pirogov, L. 1999, \aap, 348, 600
\reference{plu97} Plume, R., Jaffe, D.T., Evans, N.J. II, \etal 1997, 
ApJ, 476, 730
\reference{pcc04} Popescu, C.C., Tuffs, R.J., Kylafis, N.D., Madore, B.F. 
2004, \aap, 414, 45
\reference{Rad91} Radford, S.J.E., Delannoy, J., Downes, D., \etal 1991, 
in IAU Symp~146, 
{\it Dynamics of Galaxies and their Molecular Cloud Distributions}, 
ed. F. Combes, \& F. Casoli (Dordrecht: Kluwer), 303
\reference{Rad94} Radford, S.J.E. 1994, in {\it The Cold Universe}, ed. 
T. Montmerle, C.J. Lada, I.F. Mirabel, J. Tran Thanh Van, 
(Gif-sun-Yuette: Ed Frontieres) p. 369
\reference{rd97} Reynaud, D. \& Downes, D. 1997, \aap, 319, 737
\reference{Ric93} Rice, W. 1993, AJ, 105, 67
\reference{Ric88} Rice, W., \etal 1988, ApJS, 68, 91
\reference{rrm00} Rowan-Robinson, M. 2000, MNRAS, 316, 885
\reference{row99} Rownd, B.K., \& Young, J.S. 1999, AJ, 118, 670
\reference{Sak99} Sakamoto, K., \etal 1999, ApJ, 514, 68
\reference{San88} Sanders, D.B., Soifer, B.T., Elias, J.H., Madore, B.F.,
Matthews, K., Neugebauer, G., \& Scoville, N.Z. 1988, ApJ, 325, 74 
\reference{San89} Sanders, D.B., Phinney, E.S., Neugebauer, G., 
Soifer, B.T., \& Matthews, K. 1989, ApJ, 347, 29
\reference{San91} Sanders, D.B., Scoville, N.Z., \& Soifer, B.T. 1991, ApJ,
 370, 158
\reference{San96} Sanders, D.B., \& Mirabel, I.F. 1996, ARA\&A, 34, 749
\reference{} Sanders, D.B. 1999, astro-ph/9908297
\reference{} Schmidt, M. 1959, ApJ, 129, 243
\reference{Sco83} Scoville, N.Z., \& Young, J.S. 1983, ApJ, 265, 148
\reference{Sco97} Scoville, N.Z., Yun, M.S., \& Bryant, P.M. 1997, ApJ, 484, 702
\reference{Sco00} Scoville, N.Z., Evans, A.S., Thompson, R. \etal 2000, 
AJ, 119, 991
\reference{Shi03} Shibatsuka, T., Matsushita, S., Kohno, K., \& Kawabe, R. 
2003, PASJ, 55, 87
\reference{Soi89} Soifer, B.T. \etal 1989, \aj, 98, 766
\reference{Soi01} Soifer, B.T., Neugebauer, G., Matthews, K. \etal 2001, \aj, 122, 1213
\reference{Sol79} Solomon, P.M., Sanders, D.B., \& Scoville, N.Z. 1979,  
 in {\it  The Large Scale Characteristics of the Galaxy}, IAUS 84, 
(Dordrecht D Reidel) 35
\reference{Sol87} Solomon, P.M., Rivolo, A.R., Barrett, J., \& Yahil, A.
 1987, ApJ, 319, 730
\reference{Sol91} Solomon, P.M., \& Barrett, J.W. 1991, in IAU Symp~146, 
{\it Dynamics of Galaxies and their Molecular Cloud Distributions}, 
ed. F. Combes, \& F. Casoli (Dordrecht: Kluwer), 235
\reference{Sol90} Solomon, P.M., Downes, D., \& Radford, S.J.E. 1992, ApJ, 
387, L55
\reference{Sol92} Solomon, P.M., Radford, S.J.E., \& Downes, D. 1990, ApJ,
348, L53 
\reference{Sol97} Solomon, P.M., Downes, D., Radford, S.J.E., \& Barrett, J.W.
1997, ApJ, 478, 144 
\reference{Sol88} Solomon, P.M., \& Sage, L.J. 1988, ApJ, 334, 613
\reference{Sol03} Solomon, P.M., Vanden Bout, P., Carilli, C., \& Guelin, M. 
2003, Nature, 426, 636
\reference{} Sorai, K., Nakai, N., Kuno, N., \& Nishiyama, K. 2002, PASJ, 54,
179
\reference{Stu88} Stutzki, J., Genzel, R., Harris, A.I., Herman, J., \& Jaffe,
D.T. 1988, 330, L125
\reference{Sur99} Surace, J.A., \& Sanders, D.B. 1999, ApJ, 512, 162
\reference{Sur98} Surace, J.A., Sanders, D.B., Vacca, W.D., Veilleux, S., 
\& Mazzarella, J.M. 1998, ApJ, 492, 116
\reference{Tel93} Telesco, C.M., Dressel, L.L., \& Wolstencroft, R.D. 1993,
ApJ, 414, 120
\reference{Tac94} Tacconi, L.J., Genzel, R., Blietz, M., Cameron, M.,
Harris, A. \& Madden, S. 1994, ApJ, 426, L77
\reference{Tac97} Tacconi, L.J., \etal 1997, ApSS, 248, 59
\reference{TH92} Turner, J.L., \& Hurt, R.L. 1992, ApJ, 384, 72
\reference{TH83} Turner, J.L., \& Ho, P.T.P. 1983, ApJ, 268, L79
\reference{Tur93} Turner, J.L., et al. 1993, ApJ, 413, L19
\reference{} Veilleux, S., Kim, D.C., \& Sanders, D.B. 2002, ApJS, 143, 315
\reference{} Veilleux, S., Sanders, D.B., \& Kim, D.C. 1999, ApJ, 522, 139 
\reference{} Wang, W.H., Lo, K.Y., Gao, Y., \& Gruendl, R.A. 2001, AJ, 122, 140
\reference{} Wong, T., \& Blitz, L. 2002, ApJ, 569, 157
\reference{Wil00} Wild, W., \& Eckart, A. 2000, A\&A, 359, 483
\reference{} Xu, C., Gao, Y., Mazzarella, J., Lu, N.Y., Sulentic, J.W., \&
Domingue, D.L. 2000, ApJ, 541, 644
\reference{You86} Young, J.S., Schloerb, F.P., Kenney, J.D.P., Lord, S.D. 
1986, ApJ, 304, 443
\reference{You89} Young, J.S., Xie, S., Kenney, J.D.P., Rice, W.L. 1989, ApJS,
70, 699
\reference{You91} Young, J.S., \& Scoville, N.Z. 1991, ARA\&A, 29, 581
\reference{You99} Young, J.S., 1999, ApJ, 514, L87
\end{references}
\end{document}